\documentclass[aps,prd,amssymb,eqsecnum,nofootinbib]{revtex4}
\usepackage{amsmath}
\usepackage{amssymb}
\usepackage{epsfig}
\begin{document}

\newcommand{\lvepsilon}{\epsilon}
\newcommand{\calO}{{O}}
\title{Quasi-local contribution to the gravitational self-force}
\author{Warren G. Anderson}
\affiliation{Department of Physics, University of Wisconsin --
Milwaukee, P.O. Box 413, Milwaukee, Wisconsin, 53201, U.S.A.}
\author{\'{E}anna \'{E}. Flanagan}
\affiliation{Newman Laboratory, Cornell University, Ithaca, New York 
14853-5001, U.S.A.}
\author{Adrian C. Ottewill}
\affiliation{Department of Mathematical Physics, University College Dublin,
Belfield, Dublin 4, Ireland}

\begin{abstract}%
\vspace{0.5cm}

The gravitational self-force on a point particle moving in a vacuum background
spacetime can be expressed as an integral over the past worldline of the
particle, the so-called tail term. In this paper, we consider that piece of
the self-force obtained by integrating over a portion of the past worldline
that extends a proper time ${\Delta}{\tau}$ into the past, provided that
${\Delta}{\tau}$ does not extend beyond the normal neighborhood of the
particle. We express this ``quasi-local'' piece as a power series in the
proper time interval ${\Delta}{\tau}$.  We argue from symmetries and
dimensional considerations that the $O({\Delta}{\tau}^0)$ and
$O({\Delta}{\tau})$ terms in this power series must vanish, and compute the
first two non-vanishing terms which occur at $O({\Delta}{\tau}^2)$ and
$O({\Delta}{\tau}^3)$.  The coefficients in the expansion depend only on the
particle's four velocity and on the Weyl tensor and its derivatives at the
particle's location.  The result may be useful as a foundation for a practical
computational method for gravitational self-forces in the Kerr spacetime, in
which the portion of the tail integral in the distant past is computed
numerically from a mode sum decomposition.

\end{abstract}
\maketitle

\section{Introduction and Summary} 
\label{s:intro}

One of the outstanding open problems of classical general relativity is the
calculation of the gravitational self-force experienced by a massive particle
moving in a curved background spacetime.  Here by particle we do not mean a
point particle, but rather an extended object whose internal structure has a
negligible effect on its ``center-of-mass'' motion.  Such a particle will not
follow a geodesic of the background space-time, but rather a geodesic of the
total spacetime, whose curvature reflects both the background and the
mass/energy of the particle itself. However, if the mass of the particle is
much less than the natural length-scale of the background (i.e. the
square-root of the inverse of the curvature scale of the background), the
deviation of the particle's trajectory from a background geodesic will be small
over time-scales that are less than the natural length-scale of the
background. In this case, one can treat the difference between the total
spacetime metric and the background spacetime metric to linear order as spin-2
field generated by the particle and living on the background spacetime. This
field couples to the particle, and in this picture, it is understood as
causing a force, the \textit{self-force}, which causes the particle to deviate
from the background geodesic.

With this description, the gravitational self-force is analogous to the
self-force experienced by a electrically charged point particle coupled to a
Maxwell field in curved spacetime, or to the self-force experienced by a
particle carrying a scalar charge coupled to a massless linear scalar field in
curved spacetime.  In each of these cases, the interaction of a particle with
its own field alters its motion. In the flat spacetime limit, these forces
reduce to the familiar radiation reaction forces, as they are only associated
with accelerated motion and the resultant emission of radiation. In curved
spacetimes, however, the notion of emission of radiation can, in general, be
ambiguous. Furthermore, even in stationary spacetimes, the self-forces can
have conservative terms, as well as dissipative pieces reminiscent of
radiation reaction forces. Thus, in curved spacetimes, the notion of a
radiation reaction force is replaced by the more general notion of a
self-force.

Recently, there has been renewed interest in finding the gravitational
self-force experienced by a massive particle traveling in a curved background
spacetime. The primary impetus for this interest is the imminent construction 
of space-based gravitational wave observatories, such as LISA\cite{LISAWWW}.
One of the most interesting gravitational wave sources for these instruments
will be compact solar mass objects inspiralling into black holes of $10^3$ to 
$10^8$ solar masses out to Gpc distances\cite{FT00}. The extraction of
the maximal amount of information from such observations, however, can only 
be effected if accurate theoretical waveform templates exist. The calculation
of such templates requires precise knowledge of the orbital evolution of the
smaller object in the gravitational background of the massive black hole,
which in turn requires an accurate calculation of the gravitational
self-force.

In flat spacetime, the radiation reaction force can be thought of as the
recoil force experienced by the particle as it emits radiation.  Since the
radiation carries momentum, and since the radiation field is not generally
spherically symmetric, there is a non-zero net momentum transferred to the
particle.  This change of momentum corresponds to a force, which for an
electrically charged particle in flat spacetime is given by the
Abraham-Lorentz-Dirac formula \cite{Dirac38,Poisson99}.  This explanation of
the radiation reaction force is, of course, highly simplified. It is not
straightforward to relate the force directly to the radiation --- the force is
local to the particle while radiation is a far-field phenomenon. Nonetheless,
this simplified picture can provide intuitive insight into the phenomenon.

In curved spacetimes, qualitatively new types of self interactions occur due
to the failure of Huygens' principle (in its modern incarnation) to hold in
most geometries \footnote{This statement is in fact a modification of a
conjecture by Hadamard\cite{Hadamard23} that the only hyperbolic differential
operators whose solutions obey Huygens' principle are conformally related to
the the ordinary wave operator in Minkowski space with even numbers of
spacetime dimensions. Hadamard originally formulated this conjecture with
scalar operators in mind, but it has since been trivially extended to
operators for fields with higher spin. Counter examples to Hadamard's
conjecture have been found, originally in $\ge 6$ (but even) dimensions by
Stellmacher\cite{Stellmacher53} and later in the more physically relevant case
of $\ge 4$ (even) dimensions by G\"unther\cite{Gunther65}.  Nonetheless, these
counter examples are believed (but not yet proved) to be isolated cases. This
modified Hadamard conjecture has been proved in broad classes of spacetimes
(e.g. some algebraically special spacetimes) and work continues in this
area.}. More pedantically, it is due to the fact that solutions to wave
equations on most curved manifolds depend not only on Cauchy data directly
intersected by the  past light cone, but also on Cauchy data interior to that
intersection. The portion of the field that propagates in the null directions
along the characteristics is called the \emph{direct} part. The portion of the
field which propagates in the timelike directions in the interior of the light
cone is called the \emph{tail}. Clearly, the tail part of a particle's field
can interact with the particle, leading to a contribution to the self-force.

A general expression for the self-force on an electrically charged particle in
a curved spacetime was obtained in the seminal paper by DeWitt and
Brehme\cite{DB60} \footnote{The expression obtained by DeWitt and Brehme is
missing a term due to a trivial calculational error; see Hobbs\cite{Hobbs68}
for the correction.}.  More recently, similar expressions have been obtained
for the gravitational self-force by Mino, Sasaki and Tanaka \cite{MST97} and
by Quinn and Wald \cite{QW97}, which we review in Sec.\ \ref{s:gsf} below, and
for the scalar self-force by Quinn\cite{Quinn00}.  These results have resolved
many of the issues of principle in computing self-forces in curved spacetime.
See Ref.\ \cite{Poisson03} for a detailed review of and simplified versions of
these computations.

However, for applications to gravitational wave observations, one
needs to translate the formal expressions of Refs.\
\cite{MST97,QW97} into practical computational schemes
for computing orbits of particles in the Kerr spacetime.  The
expressions for the self-force involve the retarded Green's function
for the wave equation, and the standard method of computing this Green's
function is to use a decomposition of the field into modes.  This
mode decomposition method combines together the tail and singular
pieces of the fields in a manner that is difficult to disentangle, and
it is the tail piece of the self-field that determines the self-force.
Thus, the results of Refs.\ \cite{DB60,MST97,QW97,Quinn00} do not
directly give a simple method of computing self-forces in black hole
spacetimes.

Nonetheless, some progress has been made in calculating
the self-force for particular particle trajectories in particular spacetimes.
For specific geometries, and also in the weak-field
approximation, it has been possible to compute the tail of the Green's
function.  Some classic results have been obtained for scalar or 
electric charges in static, radial, or circular trajectories about black holes
or cosmic strings \cite{DD64, SW80, LL82, Lohiya82, ZF82,
LL84, Linet86, Smith90, AKO95, Wiseman00, PP00}.  In Schwarzschild, and for
circular or equatorial orbits in Kerr, the time-averaged non-conservative
(i.e. radiation reaction) contributions to the gravitational self-force may be
deduced using energy and angular-momentum balance arguments involving
the flux of radiation to infinity and down the black
hole\cite{TSSTN93, Shibata93, CKP94, Ryan96, KO98, Kennefick98,
Hughes00, OT00}. Furthermore, in the weak-field, slow-motion limit of
general relativity, one may use the post-Newtonian expansion to obtain
the gravitational self-force \cite{BF00, JS00, PW00}, and the result
agrees with that obtained by specializing the formal results of Refs.\
\cite{MST97,QW97} to weak fields \cite{PP00}.

While these results are encouraging, it is important and desirable to have a
framework in which arbitrary motions in black hole spacetimes can be
computed.  Recently considerable progress has been made in developing
practical computational schemes for obtaining self-forces
\footnote{See the proceedings of the Capra Ranch meetings on radiation
reaction at \texttt{http://www.lsc-group.phys.uwm.edu/\~\,patrick/ireland99/},
\texttt{http://www.tapir.caltech.edu/\~\,capra3/},
\texttt{http://www.aei-potsdam.mpg.de/lousto/capra/},
\texttt{http://cgwp.gravity.psu.edu/events/Capra5/capra5-BKP\_2002-05-24-1200.shtml}
and \texttt{http://cgwa.phys.utb.edu/Events/agendaView.php?EventID\=3}.}.
Most of these schemes are based on computing the full retarded field, which is
infinite on the particle's worldline, and regularizing it in some way to
effect the subtraction of the direct part of the field, leaving the desired
tail part.   Barack and Ori have derived a mode sum regularization
scheme\cite{Barack100, Barack101, OB00} that has been successfully applied  in
a number of cases\cite{OB00, BB00, Burko100, Burko200, BL01, BLS01}.  The
regularization parameters for this scheme have been derived from the
fundamental MSTQW equation of motion for Schwarzschild in Refs.\
\cite{BMNOS02,BO02a} and for Kerr in Ref.\ \cite{BO2003}.  Mino, Nakano and
Sasaki have developed two regularization schemes, one of which is a
mode-by-mode regularization and the second of which they dub the ``power
expansion regularization'' which involves a Post-Newtonian expansion of the
Green's function\cite{Mino98,MN98,MNS01}. Both their methods have been applied
\cite{NS01,NMS01}.  Another scheme is that of Lousto, who used zeta-function
regularization of modes for a radially infalling scalar particle in
Schwarzschild\cite{Lousto00}.

A qualitatively different method for computing self-forces in black hole
spacetimes has been suggested by Poisson and Wiseman \cite{Wiseman97}.  It is
based on a direct computation of the tail field, rather than a regularization
of the total retarded field.  The tail field can be expressed as an integral
over the past worldline of the particle.  The idea is to split this integral
into two pieces, a piece that extends back into the past a proper time
interval ${\Delta}{\tau}$, which we call the {\it quasi-local} piece, and the
remainder of the integral.  The second piece, from the distant past, can be
computed using standard multipolar decomposition of the full retarded field;
no difficulties involving divergences occur here, and thus no regularization
is needed.  The first, quasi-local piece can be computed approximately as a
power series expansion in ${\Delta}{\tau}$.

In this paper, we compute the expansion in powers of ${\Delta}{\tau}$ of the
quasi-local piece of the gravitational self force for an arbitrary vacuum
spacetime, to the first two non-trivial orders in ${\Delta}{\tau}$.  Our
result is given in Eq.\ (\ref{eq:finalans}) below, and may be useful as a
foundation for the Poisson-Wiseman scheme.  Alternatively it may be useful as
a check of numerical codes that use some regularization scheme.  At the core
of our analysis is a local expansion of the tail piece of the Green's function
for linearized perturbations.  Such local expansions of Green's functions can
be found in the literature on quantum field theory in curved spacetime; see
Refs.\ \cite{BO83,BO86,BF86} for the scalar case, Ref.\ \cite{BO86} for the
electromagnetic case, and Ref.\ \cite{AFO} for the gravitational case.  We
extend the expansion of Ref.\ \cite{AFO} to one higher order, and apply the
result to compute the quasi-local piece of the gravitational self-force.  

The organization of this paper is as follows.   In Section \ref{s:gsf} we
review the formal expression for the gravitational self-force obtained
by Mino, Sasaki and Tanaka\cite{MST97} and by Quinn and
Wald\cite{QW97}, and define the quasi-local contribution to the
self-force.  In Appendix \ref{sec:Counting}, we use symmetry and
dimensional arguments to deduce the possible terms that can appear in
the expansion of the quasi-local contribution, thereby reducing the
computation to obtaining one universal numerical coefficient at the 
leading nontrivial order [$O({\Delta}{\tau}^2)$] and four universal numerical
coefficients at the next higher order [$O({\Delta}{\tau}^3)$].  
Section \ref{sec:SFExpand} computes those numerical coefficients; the final result is given in Eq.\ (\ref{eq:finalans}). Some of the details of the computation are relegated to Appendices \ref{sec:foundations}, \ref{sec:Vexpand}, \ref{sec:Vcoeffs} and \ref{s:DetIDs}. In Section \ref{s:cases} we calculate the application of the general expression to some interesting cases. Finally, we make some concluding remarks in Section \ref{sec:conc}.

Throughout this paper we use geometrized units in which $G = c =1$,
and we adopt the sign conventions of Ref. \cite{MTW}.

\section{The Gravitational Self-Force}
\label{s:gsf}

\subsection{The Mino-Sasaki-Tanaka-Quinn-Wald formula}

Consider a point particle of mass ${\mu}$ moving on a geodesic $x^{\alpha}({\tau})$
of a background spacetime ($M,g_{{\beta}{\gamma}}$), parameterized by
proper time ${\tau}$.
Throughout this paper we assume that the background spacetime
satisfies the vacuum Einstein equation 
$
R_{{\alpha}{\beta}}=0.
$
The particle will
perturb the background geometry to linear order in ${\mu}$.  
We denote the linearized metric perturbation 
by $h_{{\alpha}{\beta}}$, and the more convenient trace-reversed form of this 
perturbation by
\begin{equation}
   {\psi}_{{\beta}{\gamma}} \equiv h_{{\beta}{\gamma}}-\frac{1}{2} ~g_{{\beta}{\gamma}} ~
      h_{{\mu}{\nu}}~g^{{\mu}{\nu}}.
\label{eq:TRmetric}
\end{equation}
We raise and lower indices with the background
metric.  We specialize throughout to the Lorentz gauge defined by
\begin{equation}
{\psi}^{{\beta}{\gamma}}{_{;{\gamma}}} =0,
\label{eq:Lorentz}
\end{equation} 
where the semicolon denotes a covariant derivative with respect to the
background metric $g_{{\alpha}{\beta}}$.  
In this gauge the linearized Einstein field equations 
take the form of the simple wave equation
\begin{equation}
   ({\Box}g_{{\mu}{\alpha}}g_{{\nu}{\beta}}+2C_{{\mu}{\alpha}{\nu}{\beta}}){\psi}^{{\mu}{\nu}}
      = - 16 {\pi}T_{{\alpha}{\beta}},
\label{eq:LEFE}
\end{equation}
where ${\Box}$ and $C_{{\mu}{\alpha}{\nu}{\beta}}$ are the D'Alembertian and Weyl
tensor associated with the background metric $g_{{\beta}{\gamma}}$,
and $T_{{\alpha}{\beta}}$ is the linearized stress energy tensor.  The
wave equation (\ref{eq:LEFE}) can be 
solved using the retarded Green's function
$G_{\text{ret}}^{{\mu}{\nu}{{\alpha}'}{{\beta}'}}$, which is defined by the
equation 
\begin{equation}
   ({\Box}g_{{\mu}{\alpha}}g_{{\nu}{\beta}}+2C_{{\mu}{\alpha}{\nu}{\beta}})
      G_{\text{ret}}^{{\mu}{\nu}{{\alpha}'} {{\beta}'}}(x,x')=-g_{({\alpha}}{^{{\alpha}'}}
      g_{{\beta})}{^{{\beta}'}} {\delta}^4(x,x'),
\label{eq:LGFE}
\end{equation}
and by the fact that it has support only when $x'$ is in the causal past of
$x$.  Here $g_{\alpha}{^{{\alpha}'}}$ is the parallel displacement bivector
\cite{DB60,C76,C78}, and  
${\delta}^4(x,x') = {\delta}^4(x-x^{\prime})/\sqrt{-g}$ is a generalized
Dirac delta function\footnote{Note that because we have
used the sign convention of Misner, Thorne and Wheeler \protect{\cite{MTW}},   
our Weyl tensor has the opposite sign to that of Mino, Sasaki and
Tanaka \cite{MST97}. Note also that our Green's function is defined to
be one half that of Ref.\ \protect{\cite{MST97}}.}.
The retarded solution to Eq.\ (\ref{eq:LEFE}) can be written in terms
of the Green's function as
\begin{equation}
   {\psi}^{{\mu}{\nu}}_{\rm ret}(x)=16{\pi}{\int}d^4x^{\prime}\, \sqrt{-g(x^{\prime})} \,  
      G_{\text{ret}}^{{\mu}{\nu}}{_{{\alpha}'{\beta}'}}(x,x')T^{{\alpha}'{\beta}'}(x'). 
\label{eq:pertint}
\end{equation} 
For the point particle source, the stress energy tensor is given by
\begin{equation}
   T_{{\alpha}{\beta}}(x)={\mu}{\int}_{-{\infty}}^{\infty}\,{\delta}^4[x,x'({\tau}')] u_{\alpha}({\tau}') 
   u_{\beta}({\tau}') d{\tau}',
\label{eq:SET}
\end{equation} 
where $u_{\alpha}({\tau})$ is the particle's four velocity, and inserting
this into Eq.\ (\ref{eq:pertint}) gives 
\begin{equation}
   {\psi}^{{\mu}{\nu}}_{\rm ret}(x)=16{\pi}{\mu}{\int}_{-{\infty}}^{\infty}d{\tau}^{\prime}\,
      G_{\text{ret}}^{{\mu}{\nu}}{_{{\alpha}'{\beta}'}}[x,x^{\prime}({\tau}^{\prime})]
      \, u^{{\alpha}^{\prime}}({\tau}^{\prime}) u^{{\beta}^{\prime}}({\tau}^{\prime}).
\label{eq:pertint1}
\end{equation} 

Now one would expect the particle to move on a geodesic of the total
metric $g_{{\alpha}{\beta}} + {\psi}^{\rm ret}_{{\alpha}{\beta}} - 
g_{{\alpha}{\beta}} g^{{\gamma}{\delta}} {\psi}^{\rm ret}_{{\gamma}{\delta}}/2$.
Such geodesic motion  
would be equivalent to motion for which the mass times acceleration
with respect to $g_{{\alpha}{\beta}}$ is
\begin{equation}
   f^{\alpha}= {\mu}\, P^{{\alpha}{\beta}{\gamma}{\delta}}\,
   {\psi}^{\text{ret}}_{{\beta}{\gamma};{\delta}},
\label{eq:selfforce naive}
\end{equation}
where the tensor $P_{{\alpha}{\beta}{\gamma}{\delta}}$ is given by
\begin{equation}
   P^{{\alpha}{\beta}{\gamma}{\delta}}\equiv-\frac{1}{2}u^{\alpha}u^{\beta}u^{\gamma}u^{\delta}-
      g^{{\alpha}({\beta}} u^{{\gamma})} u^{\delta}+\frac{1}{2}g^{{\alpha}{\delta}} u^{\beta}
      u^{\gamma}+ \frac{1}{4}u^{\alpha}g^{{\beta}{\gamma}} u^{\delta}+ \frac{1}{4}
      g^{{\alpha}{\delta}} g^{{\beta}{\gamma}}.
\label{eq:Pdefn}
\end{equation}
However, the retarded field ${\psi}^{\rm ret}_{{\mu}{\nu}}$ and its gradient
${\psi}^{\rm ret}_{{\mu}{\nu};{\lambda}}$ are divergent on the particle's
worldline, so the naive expression (\ref{eq:selfforce naive}) for the self
force is ill defined.  Instead, the correct expression for the self
force is given by Eq.\ (\ref{eq:selfforce naive}) with the retarded
field replaced by the so called tail field ${\psi}^{\rm tail}_{{\mu}{\nu}}$
\cite{MST97,QW97}: 
\begin{equation}
   f^{\alpha}= {\mu}\, P^{{\alpha}{\beta}{\gamma}{\delta}}\,
   \left< {\psi}^{\text{tail}}_{{\beta}{\gamma};{\delta}} \right>.
\label{eq:selfforce}
\end{equation}
Here the angular brackets $\left< \ldots \right>$ denote the result
obtained by averaging over a 2 sphere of some small radius $r$ about the
particle, in the spatial hypersurface orthogonal to $u^{\alpha}$, and by
taking the limit $r \to 0$.
The tail field ${\psi}^{\rm tail}_{{\mu}{\nu}}(x)$ is defined by truncating
the integral (\ref{eq:pertint1}) over the particle's worldline to
exclude the contribution from the direct part of the Green's function:
\begin{equation}
   {\psi}^{{\mu}{\nu}}_{\rm tail}(x)=16{\pi}{\mu}{\displaystyle 
      \lim_{{\epsilon}\to 0^+} }
      \, {\int}_{-{\infty}}^{{\tau}_{\rm ret}(x)-{\epsilon}} d{\tau}^{\prime}\, 
      G_{\text{ret}}^{{\mu}{\nu}}{_{{\alpha}'{\beta}'}}[x,x^{\prime}({\tau}^{\prime})]
      \, u^{{\alpha}^{\prime}}({\tau}^{\prime}) u^{{\beta}^{\prime}}({\tau}^{\prime}).
\label{eq:taildef}
\end{equation} 
Here ${\tau}_{\rm ret}(x)$ is the value of proper time ${\tau}$ at the
point where the worldline intersects the past lightcone of the point $x$.  
The remaining, direct portion of the field is 
\begin{equation}
   {\psi}^{{\mu}{\nu}}_{\rm direct}(x)=   {\psi}^{{\mu}{\nu}}_{\rm ret}(x) - 
   {\psi}^{{\mu}{\nu}}_{\rm tail}(x)=
16{\pi}{\mu}{\displaystyle 
      \lim_{{\epsilon}\to 0^+} }
      \, {\int}_{{\tau}_{\rm ret}(x)-{\epsilon}}^{{\tau}_{\rm ret}(x)+{\epsilon}} 
d{\tau}^{\prime}\,
      G_{\text{ret}}^{{\mu}{\nu}}{_{{\alpha}'{\beta}'}}[x,x^{\prime}({\tau}^{\prime})] 
      \, u^{{\alpha}^{\prime}}({\tau}^{\prime}) u^{{\beta}^{\prime}}({\tau}^{\prime}).
\label{eq:pertint1a}
\end{equation} 

If we now take the gradient of the tail field (\ref{eq:taildef}) in order
to substitute  
into the formula (\ref{eq:selfforce}) for the self-force, we obtain
two terms: a term generated by the action of the gradient operator on
the quantity ${\tau}_{\rm ret}(x)$, and a term generated by the action
of the gradient operator on the retarded Green's function.  
The first contribution gives an expression which has a direction
dependent limit on the world line, but which vanishes once the average
$\left< \ldots \right>$ is taken \cite{QW97,DW03}.  The second
contribution is continuous on the worldline (so the averaging can be
dispensed with), and yields for the self-force the expression
\cite{MST97,QW97}
\begin{equation}
f^{\alpha}({\tau}) = 
16{\pi}{\mu}^2 P^{{\alpha}{\beta}{\gamma}{\delta}}\,{\displaystyle 
      \lim_{{\epsilon}\to 0^+} }
      \, {\int}_{-{\infty}}^{{\tau}-{\epsilon}} d{\tau}^{\prime}\, 
      G^{\text{ret}}_{{\beta}{\gamma}{\beta}^{\prime}{\gamma}^{\prime};{\delta}}[x,x^{\prime}({\tau}^{\prime})] 
      \, u^{{\beta}^{\prime}}({\tau}^{\prime}) u^{{\gamma}^{\prime}}({\tau}^{\prime}).
\label{eq:self force final}
\end{equation}

As an aside, we note that Detweiler and Whiting \cite{DW03} have introduced an
alternative splitting of the retarded field of the form
\begin{equation}
{\psi}_{\rm ret}^{{\mu}{\nu}} = {\psi}_{\rm sing}^{{\mu}{\nu}} + {\psi}_{\rm
  regular}^{{\mu}{\nu}},
\end{equation}
where the singular piece ${\psi}_{\rm sing}^{{\mu}{\nu}}$ is a solution of
the inhomogeneous wave equation (\ref{eq:LEFE}) which can be
computed locally, and ${\psi}_{\rm regular}^{{\mu}{\nu}}$ is a solution of
the corresponding homogeneous wave equation such that the self force
is given by the expression (\ref{eq:selfforce naive}) with ${\psi}_{\rm
ret}^{{\mu}{\nu}}$ replaced by ${\psi}_{\rm regular}^{{\mu}{\nu}}$.  This
alternative formulation also gives rise to the final formula (\ref{eq:self
force final}) for the self force.

\subsection{Equation of motion}

The self-force formula (\ref{eq:self force final}) discussed above
was defined only for a geodesic worldline.  Therefore it is necessary
to supplement the self-force formula 
with a prescription for computing
the motion of a point particle that includes the effect of 
the self force to leading order in the particle's mass ${\mu}$.  
The reason that finding such a prescription is not entirely trivial is
the following \cite{MST97,QW97}:  the linearized Einstein equation
admits solutions only if its source, the particle's stress energy tensor
(\ref{eq:SET}), is conserved.  However, the stress energy tensor for a
point particle is conserved only if the worldline is a
geodesic.  Therefore it is not straightforward to define the
self-force on a non-geodesic worldline.  

Quinn and Wald \cite{QW97} suggested the following method of resolving
this difficulty.  They define a self-force for a non-geodesic worldline
by relaxing the Lorentz gauge condition (\ref{eq:Lorentz}) while
retaining the form (\ref{eq:LEFE}) of the linearized Einstein
equation.  The justification for relaxing the gauge condition is that
the associated errors are quadratic in the mass ${\mu}$, while the
self-acceleration is linear in ${\mu}$.  Their equation of motion is then
\begin{equation}
{\mu}a^{\alpha}= f^{\alpha},
\label{eq:eomQW}
\end{equation}
with $f^{\alpha}$ given by Eq.\ (\ref{eq:self force final}), modified as
described above, and $a^{\alpha}$ is the 4-acceleration with respect to
the background metric.  The equation of motion (\ref{eq:eomQW}) is an
integro-differential equation which is non-local in time.  

Here, we suggest an alternative equation of motion, which gives the
same results as Eq.\ (\ref{eq:eomQW}) to leading order in ${\mu}$.
First, for any point ${\cal 
P}$ in spacetime, and for any unit, future-directed timelike vector ${\vec 
u}$ at ${\cal P}$, we define the self-force vector $f^{\alpha}=
f^{\alpha}({\cal P}, {\vec u})$ to be self force obtained from the 
prescription (\ref{eq:self force final}) for the particular geodesic
which extends
into the past from ${\cal P}$ whose tangent at ${\cal P}$ is ${\vec
u}$.  The equation of motion for the perturbed worldline
$x^{\alpha}({\tau})$ is then
\begin{equation}
{\mu}a^{\alpha}= f^{\alpha}[{\vec x}({\tau}),{\vec u}({\tau})].
\label{eq:eomUS}
\end{equation}
In other words, the worldline is such that its acceleration at any
point is the acceleration that is obtained from the integral
(\ref{eq:self force final}) for the geodesic which is tangent to the
worldline at that point.  The justification for using the
instantaneously-tangential 
geodesic rather than the true worldline is essentially the reduction
of order argument discussed in Refs.\ \cite{FW96,QW97}.
The equation of motion (\ref{eq:eomUS}) is a second order
differential equation which is local in time.

\subsection{Definition of the quasi-local piece of the self-force}
\label{sec:quasilocal}

In this paper we evaluate not the entire integral
over the worldline in the self-force formula (\ref{eq:self force
final}), but instead that portion of that integral near the
particle with proper times ${\tau}^{\prime}$ in the range ${\tau}- {\Delta}
{\tau}\le {\tau}^{\prime}\le {\tau}$.  Specifically, we define 
\begin{equation}
f^{\alpha}_{\rm QL}({\tau},{\Delta}{\tau}) = 
16{\pi}{\mu}^2 P^{{\alpha}{\beta}{\gamma}{\delta}}\,{\displaystyle 
      \lim_{{\epsilon}\to 0^+} }
      \, {\int}_{{\tau}-{\Delta}{\tau}}^{{\tau}-{\epsilon}} d{\tau}^{\prime}\, 
      G^{\text{ret}}_{{\beta}{\gamma}{\beta}^{\prime}{\gamma}^{\prime};{\delta}}[x,x^{\prime}({\tau}^{\prime})] 
      \, u^{{\beta}^{\prime}}({\tau}^{\prime}) u^{{\gamma}^{\prime}}({\tau}^{\prime}).
\label{eq:QLdef}
\end{equation}
Here the subscript ``QL'' denotes ``quasi-local''.  By comparing with
Eq.\ (\ref{eq:self force final}) we see that the entire self-force is
obtained from $f_{\rm QL}^{\alpha}({\tau},{\Delta}{\tau})$ in the limit where
${\Delta}{\tau}\to {\infty}$.  We will derive an approximate power
series expansion of $f_{\rm QL}^{\alpha}({\tau},{\Delta}{\tau})$ that is
valid in the limit ${\Delta}{\tau}\to 0$.

Within a sufficiently small neighborhood of the point $x^{\alpha}=
x^{\alpha}({\tau})$, the retarded Green's  
function can be written in the Hadamard form
\begin{equation}
   G_{\rm ret}^{{\mu}{\nu}{{\alpha}'} {{\beta}'}}(x,x')=
         \frac{{\Theta}[\Sigma(x,x')]}{4{\pi}}
      {\times}\Bigl\{U^{{\mu}{\nu}{{\alpha}'}{{\beta}'}}(x,x')~
         {\delta}[{\sigma}(x,x')]
      -V^{{\mu}{\nu}{{\alpha}'}{{\beta}'}}(x,x')~
         {\Theta}[-{\sigma}(x,x')]\Bigr\}.
\label{eq:Hadamard}
\end{equation}
Here $\Sigma(x,x')$ is an arbitrary function which is positive for $x$ in the
causal future of $x'$ and negative otherwise, ${\Theta}[{\cdot}]$ is the Heaviside 
step function, ${\sigma}(x,x')$ is Synge's world function
\cite{Synge,Poisson03}, ${\delta}[{\cdot}]$ is the ordinary Dirac   
delta distribution, and $U^{{\mu}{\nu}{{\alpha}'}{{\beta}'}}(x,x')$ and
$V^{{\mu}{\nu}{{\alpha}'}{{\beta}'}}(x,x')$ are smooth functions.
The part of the Green's function proportional to
${\delta}({\sigma})$ is called the {\it direct part}, and the part of
the Green's function proportional
to the function $V^{{\mu}{\nu}{{\alpha}'}{{\beta}'}}$ as the {\it tail part}.
For sufficiently small ${\Delta}{\tau}$, the portion of the worldline
between $x^{\alpha}({\tau}-{\Delta}{\tau})$ and $x^{\alpha}({\tau})$ 
that arises in Eq.\ (\ref{eq:QLdef}) will lie
inside the neighborhood where the Hadamard expression
(\ref{eq:Hadamard}) is valid.  Inserting this expression into the
formula (\ref{eq:QLdef}) for $f_{\rm QL}^{\alpha}$ gives
\begin{equation}
   f^{\alpha}_{\rm QL}({\tau},{\Delta}{\tau}) =-
   4\,{\mu}^2\,P^{{\alpha}{\beta}{\gamma}{\delta}}\, 
   {\int}_{{\tau}- {\Delta}{\tau}}^{\tau}
   V_{{\beta}{\gamma}{\beta}'{\gamma}';{\delta}}[x,x'({\tau}')] u^{{\beta}'}({\tau}') 
   u^{{\gamma}'}({\tau}') d{\tau}'.
\label{eq:QLint}
\end{equation}
Note that the direct part of the Green's function 
does not contribute to the expression (\ref{eq:QLint}), because of
the limiting process involving ${\epsilon}$ in Eq.\ (\ref{eq:QLdef}).
That limiting process is unnecessary for the tail contribution which
gives an integrand that is finite at ${\tau}^{\prime}={\tau}$; hence there
is no limiting 
process in the final result (\ref{eq:QLint}).

The integral (\ref{eq:QLint}) can be approximately evaluated for
small ${\Delta}{\tau}$ using a covariant local expansion of 
$V_{{\beta}{\gamma}{\beta}'{\gamma}'}(x,x')$ in a neighborhood of $x^{\prime}=x$.
The result is of the form
\begin{equation}
   f_{QL\,{\alpha}}({\tau},{\Delta}
   {\tau})=f^{(0)}_{\alpha}({\tau})+f^{(1)}_{\alpha}({\tau}) {\Delta}
   {\tau}+f^{(2)}_{\alpha}({\tau}) {\Delta}{\tau}^2+f^{(3)}_{\alpha}({\tau})
   {\Delta}{\tau}^3+{\calO}({\Delta}{\tau}^4). 
\label{eq:forceexpand}
\end{equation}
Because this expansion arises from a covariant Taylor series, the coefficients
$f_{\alpha}^{(j)}({\tau})$ are purely local geometric quantities evaluated
at the present position $x^{\alpha}({\tau})$ of the particle. In Appendix
\ref{sec:Counting} we derive the geometric content of these coefficients
using simple counting and dimensional arguments.  We find that
\begin{eqnarray}
\label{eq:f0ans}
f_{\alpha}^{(0)} &=& 0, \\
\label{eq:f1ans}
f_{\alpha}^{(1)} &=& 0, \\
\label{eq:f2ans}
f_{\alpha}^{(2)} &=& c_0 {\mu}^2 ({\delta}_{\alpha}^{\beta}+ u_{\alpha}u^{\beta})
C_{{\beta}{\gamma}{\delta}{\varepsilon}} C_{{\sigma}\ \, {\rho}}^{\ \, {\gamma}\, \
  {\varepsilon}} u^{\delta}u^{\sigma}u^{\rho}, \\
\label{eq:f3ans}
f_{\alpha}^{(3)} &=& {\mu}^2 ({\delta}_{\alpha}^{\beta}+ u_{\alpha}u^{\beta})
u^{\gamma}u^{\delta}\Bigg\{
c_1 C_{{\gamma}{\mu}{\delta}{\nu}}
C_{{\varepsilon}\ \, {\sigma}\ \,;{\beta}}^{\ \, {\mu}\ \, {\nu}} 
u^{\varepsilon}u^{\sigma}
+ c_2 C_{{\beta}{\gamma}{\mu}{\delta};{\nu}} C_{{\varepsilon}\ \, {\sigma}}^{\ \,
  {\mu}\ \, {\nu}} 
u^{\varepsilon}u^{\sigma}\nonumber \\
&& + c_3\left[
  \frac{1}{2} C_{{\mu}{\nu}{\gamma}{\lambda}} C^{{\mu}{\nu}\ \,{\lambda}}_{\ \ \
    {\delta}\ \,\, ;{\beta}} + 
C_{{\mu}{\varepsilon}{\gamma}{\lambda}} C^{{\mu}\ \ \, {\lambda}}_{\ \,
{\sigma}{\delta}\ \,\, ;{\beta}} u^{\varepsilon}u^{\sigma}\right] 
+c_4 \left[
  \frac{1}{2} C_{{\mu}{\nu}{\gamma}{\lambda}} C^{{\mu}{\nu}\ \ \, \,;{\lambda}}_{\ \ \
    {\delta}{\beta}} + 
C_{{\mu}{\varepsilon}{\gamma}{\lambda}} C^{{\mu}\ \ \ \ ;{\lambda}}_{\ \,
{\sigma}{\delta}{\beta}} u^{\varepsilon}u^{\sigma}\right] 
\Bigg\},
\end{eqnarray}
where $c_0, c_1, \ldots, c_4$ are dimensionless numerical
coefficients that are as yet undetermined.  
The calculation of these coefficients proves to be the most
difficult part of determining the expansion of $f^{\alpha}_{QL}$, and is the
topic of the next section.  

\section{Expansion of the self-force}
\label{sec:SFExpand}

In this section we use a local covariant expansion of the retarded
Green's function to compute the numerical coefficients appearing in
the expansion (\ref{eq:forceexpand}) of the self force.
We use the formalism of bitensors developed by DeWitt and Brehme
\cite{DB60}; see Poisson \cite{Poisson03} for a recent detailed
review of this formalism.  A fundamental role in this formalism is
played by Synge's world function ${\sigma}(x,x^{\prime})$ and its
derivative ${\sigma}_{;{\alpha}}(x,x^{\prime})$ (see Appendix
\ref{sec:foundations}).  The local covariant
expansion of any bitensor $T(x,x^{\prime})$ takes the form
\begin{equation}
\label{eq:taylor0}
T(x,x^{\prime}) = \sum_{n=0}^{\infty}\frac{1}{n!}
t_{n}^{{{\alpha}_1}\ldots{{\alpha}_n}}(x) {\sigma}_{;{\alpha}_1}(x,x^{\prime}) \ldots
{\sigma}_{;{\alpha}_n}(x,x^{\prime}),
\end{equation}
where the coefficients $t_n^{{{\alpha}_1}\ldots{{\alpha}_n}}$ are local
tensors at $x$.  For bookkeeping purposes we define $s^2 = |{\sigma}|$,
then it follows from Eq.\ (\ref{eq:identity00}) below that the $n$th
term in Eq.\ (\ref{eq:taylor0}) scales as $s^n$.  We shall use $s$ as
an expansion parameter throughout our computations.

As a foundation for the expansion of the retarded Green's function, we
compute in Appendix \ref{sec:foundations} local covariant expansions of
a number of fundamental bitensors, including the second derivative
${\sigma}_{;{\alpha}{\beta}}$ of the world function, various covariant
derivatives of the parallel displacement bivector $g_{\alpha}^{\
\,{\alpha}'}$, and the Van-Vleck Morette determinant ${\Delta}$, to order
$O(s^5)$ beyond the leading order.  These expansions were originally
computed to $O(s^4)$ by Christensen \cite{C76,C78}, and extended to
$O(s^5)$ by Brown and Ottewill \cite{BO83,BO86}.  Our results agree
with those of Brown and Ottewill, except for one case where we correct
their result [Eq.\ (\ref{eq:boxdelta0}) below].

In Appendix \ref{sec:Vexpand} we compute the expansion to order
$O(s^3)$ of tail portion $V_{{\alpha}{\beta}{\alpha}'{\beta}'}$ of the retarded Green's
function, extending previous work of 
Allen, Folacci and Ottewill \cite{AFO} who computed the expansion to
order $O(s^2)$.  The result is of the form
\begin{eqnarray}
   V_{{\alpha}{\beta}{\alpha}'{\beta}'}&=&
g_{{\alpha}'}^{\ \ {\gamma}} g_{{\beta}'}^{\ \ {\delta}} 
\bigg[
      v^0_{{\alpha}{\beta}{\gamma}{\delta}}(x)
      +v^0_{{\alpha}{\beta}{\gamma}{\delta}{\varepsilon}}(x){\sigma}^{;{\varepsilon}}
      +\frac{1}{2}v^0_{{\alpha}{\beta}{\gamma}{\delta}{\varepsilon}{\zeta}}(x)
         {\sigma}^{;{\varepsilon}}{\sigma}^{;{\zeta}}
      +\frac{1}{6}v^0_{{\alpha}{\beta}{\gamma}{\delta}{\varepsilon}{\zeta}{\eta}}(x)
         {\sigma}^{;{\varepsilon}}{\sigma}^{;{\zeta}}{\sigma}^{;{\eta}} \nonumber
      \\
\mbox{} && + 
      v^1_{{\alpha}{\beta}{\gamma}{\delta}}(x) \, {\sigma}
      +v^1_{{\alpha}{\beta}{\gamma}{\delta}{\varepsilon}}(x)\, {\sigma}\,
      {\sigma}^{;{\varepsilon}} +O(s^4)\bigg],
\label{eq:CTSEVn0}
\end{eqnarray}
{\it cf}.\ Eqs.\ (\ref{eq:Vexpand}) and
(\ref{eq:CTSEVn}) below.  Here the various expansion coefficients
$v^0_{{\alpha}\ldots {\eta}}(x)$ and $v^1_{{\alpha}\ldots {\eta}}(x)$ are given in
Eqs.\ (\protect{\ref{eq:v01}}) -- (\protect{\ref{eq:v04}})
and (\ref{eq:v11}) -- (\ref{eq:v12}) below.

We now turn to evaluation of the integrand in the expression
(\ref{eq:QLint}) for the quasi-local piece of the self force.
First, we note that the four velocity is parallel transported along
the worldline, so we can make the replacement
\begin{equation}
   V^{{\beta}{\gamma}}{_{{\beta}'{\gamma}';{\delta}}}u^{{\beta}'}u^{{\gamma}'} = 
      \left[ V^{{\beta}{\gamma}}{_{{\beta}'{\gamma}';{\delta}}}
      g^{{\beta}'}{_{\mu}}g^{{\gamma}'}{_{\nu}} \right] u^{\mu}u^{\nu}.
\label{eq:QLintegrand}
\end{equation}
We can rewrite the first factor on the right hand side as
\begin{equation}
 V^{{\beta}{\gamma}}{_{{\beta}'{\gamma}';{\delta}}}
  g^{{\beta}'}{_{\mu}}g^{{\gamma}'}{_{\nu}} = 
 {\bar V}^{{\beta}{\gamma}}_{\ \ \ \,{\mu}{\nu};{\delta}} 
+ Q_{{\mu}{\varepsilon}{\delta}} {\bar V}^{{\beta}{\gamma}{\varepsilon}}_{\ \
\ \   \,{\nu}}  
+ Q_{{\nu}{\varepsilon}{\delta}} {\bar V}^{{\beta}{\gamma}\ \,{\varepsilon}}_{\ \
\   \,{\mu}}  
\end{equation}
where we have defined ${\bar V}_{{\alpha}{\beta}{\gamma}{\delta}} =
g^{\ \,{\alpha}'}_{\gamma}g^{\ \,{\beta}'}_{\delta}V_{{\alpha}{\beta}{\alpha}'{\beta}'}$ and
we have used the definition (\ref{eq:Qdef}) of the tensor
$Q_{{\alpha}{\beta}{\gamma}}$.  
Using the expansions (\ref{eq:CTSEVn0}), (\ref{eq:sigmaexpand}),
and (\ref{eq:Qexpand}) we now obtain
\begin{eqnarray}
   V_{{\beta}{\gamma}{\beta}'{\gamma}';{\delta}}g^{{\beta}'}{_{\mu}}g^{{\gamma}'}{_{\nu}}&=&
{\cal V}_{{\beta}{\gamma}{\mu}{\nu}{\delta}}
+{\cal V}_{{\beta}{\gamma}{\mu}{\nu}{\delta}}^{\ \ \ \ \ \ \ {\rho}}
{\sigma}_{;{\rho}}
+{\cal V}_{{\beta}{\gamma}{\mu}{\nu}{\delta}}^{\ \ \ \ \ \ \ {\rho}{\eta}} 
{\sigma}_{;{\rho}} {\sigma}_{;{\eta}} + O(s^3),
\label{eq:mainexpand}
\end{eqnarray}
where
\begin{eqnarray}
\label{eq:calVdef0}
   {\cal V}_{{\beta}{\gamma}{\mu}{\nu}{\delta}}&=&
v^0_{{\beta}{\gamma}{\mu}{\nu};{\delta}}+v^0_{{\beta}{\gamma}{\mu}{\nu}{\delta}},
 \\ \mbox{} 
{\cal V}_{{\beta}{\gamma}{\mu}{\nu}{\delta}{\rho}}
&=&
\label{eq:calVdef1}
v^0_{{\beta}{\gamma}{\mu}{\nu}{\rho};{\delta}}
         +v^0_{{\beta}{\gamma}{\mu}{\nu}{\rho}{\delta}}
         +v^1_{{\beta}{\gamma}{\mu}{\nu}}g_{{\rho}{\delta}}
         -v^0_{{\beta}{\gamma}{\lambda}{\nu}}C^{\lambda}{_{{\mu}{\delta}{\rho}}},
\\
{\cal V}_{{\beta}{\gamma}{\mu}{\nu}{\delta}{\rho}{\tau}}
&=& 
-\frac{1}{3}v^0_{{\beta}{\gamma}{\mu}{\nu}{\lambda}}
            C^{\lambda}{_{{\rho}{\delta}{\tau}}}
         +\frac{1}{2}v^0_{{\beta}{\gamma}{\mu}{\nu}{\rho}{\tau};{\delta}}
         +\frac{1}{2}v^1_{{\beta}{\gamma}{\mu}{\nu};{\delta}}g_{{\rho}{\tau}}
         +\frac{1}{2}v^0_{{\beta}{\gamma}{\mu}{\nu}{\rho}{\tau}{\delta}}
         +\frac{3}{2}v^1_{{\beta}{\gamma}{\mu}{\nu}({\delta}}g_{{\rho}{\tau})}\nonumber \\
         &&\phantom{+\Bigl(}
         -v^0_{{\beta}{\gamma}{\lambda}{\nu}{\rho}}C^{\lambda}{_{{\mu}{\delta}{\tau}}}
         +\frac{1}{3}v^0_{{\beta}{\gamma}{\lambda}{\nu}}
      C^{\lambda}{_{{\mu}{\delta}{\rho};{\tau}}}.
    \label{eq:calVdef2}
\end{eqnarray}
It is understood that the right hand sides of Eqs.\
(\ref{eq:calVdef0}) -- (\ref{eq:calVdef2}) are to be symmetrized on
the index pair $({\mu}{\nu})$, and on the index pair
$({\rho}{\tau})$ if present.
We also
note that ${\sigma}^{;{\alpha}}(x,x^{\prime})$ is proportional to the
tangent to the geodesic joining $x = x^{\alpha}({\tau})$ to $x^{\prime}=
x^{{\alpha}'}({\tau}')$, {\it i.e.}, the four velocity $u^{\alpha}$.  It
follows from the normalization condition 
(\ref{eq:identity00}) that 
\begin{equation}
   {\sigma}^{;{\alpha}}(x,x')= ({\tau}- {\tau}^{\prime}) u^{\alpha},  
\label{eq:sigmatou}
\end{equation}
where $u^{\alpha}$ is the four velocity at $x^{\alpha}({\tau})$.  
Finally, 
substituting the formulae (\ref{eq:QLintegrand}),
(\ref{eq:mainexpand}) and (\ref{eq:sigmatou}) into the expression
(\ref{eq:QLint}) gives
\begin{equation}
   f^{\alpha}_{\rm QL}({\tau},{\Delta}{\tau}) =-
   4\,{\mu}^2\,P^{{\alpha}{\beta}{\gamma}{\delta}}\,u^{\mu}u^{\nu}
   {\int}_{{\tau}- {\Delta}{\tau}}^{\tau} d{\tau}' \left\{
{\cal V}_{{\beta}{\gamma}{\mu}{\nu}{\delta}}
+{\cal V}_{{\beta}{\gamma}{\mu}{\nu}{\delta}}^{\ \ \ \ \ \ \ {\rho}}
u_{\rho}({\tau}- {\tau}')
+{\cal V}_{{\beta}{\gamma}{\mu}{\nu}{\delta}}^{\ \ \ \ \ \ \ {\rho}{\tau}} 
u_{\rho} u_{\tau} ({\tau}-{\tau}')^2+ O\left[ ({\tau}- {\tau}')^3\right]
\right\}.
\label{eq:QLint2}
\end{equation}
Evaluating the integral over ${\tau}'$ gives an expansion of $f_{\rm
QL}^{\alpha}$ of the form (\ref{eq:forceexpand}), where the coefficients
are 
\begin{eqnarray}
\label{eq:f0ans1}
f^{(0)\,{\alpha}} &=& 0, \\
\label{eq:f1ans1}
f^{(1)\,{\alpha}} &=& - 4 {\mu}^2 P^{{\alpha}{\beta}{\gamma}{\delta}} u^{\mu}u^{\nu}
  \, {\cal V}_{{\beta}{\gamma}{\mu}{\nu}{\delta}}, \\
\label{eq:f2ans1}
f^{(2)\,{\alpha}} &=& - 2 {\mu}^2 P^{{\alpha}{\beta}{\gamma}{\delta}} u^{\mu}u^{\nu}
    \, {\cal V}_{{\beta}{\gamma}{\mu}{\nu}{\delta}}^{\ \ \ \ \ \ \ \ {\rho}} u_{\rho}, 
  \\ 
\label{eq:f3ans1}
f^{(3)\,{\alpha}} &=& - \frac{4}{3} {\mu}^2 P^{{\alpha}{\beta}{\gamma}{\delta}}
    u^{\mu}u^{\nu}\, {\cal V}_{{\beta}{\gamma}{\mu}{\nu}{\delta}}^{\ \ \ \ \ \ \
    \ {\rho}{\tau}} u_{\rho}u_{\tau}.
\end{eqnarray}
The various coefficients ${\cal V}_{{\alpha}\ldots {\eta}}$ are obtained
from the formulae (\ref{eq:calVdef0}) -- (\ref{eq:calVdef2}) and are
tabulated in Appendix \ref{sec:Vcoeffs}.  We now substitute those
expressions into Eqs.\ (\ref{eq:f0ans1}) -- (\ref{eq:f3ans1}) and
use the definition (\ref{eq:Pdefn}) of the projection tensor
$P^{{\alpha}{\beta}{\gamma}{\delta}}$.   After a considerable
amount of algebra we obtain coefficients $f^{(j)}_{\alpha}$ of the form
(\ref{eq:f0ans}) --  
(\ref{eq:f2ans}), as expected.  The numerical values of the
coefficients are 
\begin{equation}
c_0 = -1
\end{equation}
and 
\begin{equation}
c_1 = \frac{1}{6}, \ \ \ c_2 = - \frac{3}{20},\ \ \ c_3 =
\frac{1}{3}, \ \ \ \ c_4 = - \frac{19}{60}.
\end{equation}
Our final expression for the quasi-local piece of the self-force is
therefore
\begin{eqnarray}
   f_{QL\,{\alpha}}({\tau},{\Delta}{\tau})
&=& - {\mu}^2 ({\delta}_{\alpha}^{\beta}+ u_{\alpha}u^{\beta})
C_{{\beta}{\gamma}{\delta}{\varepsilon}} C_{{\sigma}\ \, {\rho}}^{\ \, {\gamma}\, \
  {\varepsilon}} u^{\delta}u^{\sigma}u^{\rho}{\Delta}{\tau}^2 
+ {\mu}^2 ({\delta}_{\alpha}^{\beta}+ u_{\alpha}u^{\beta})
u^{\gamma}u^{\delta}\Bigg\{
\frac{1}{6}
C_{{\gamma}{\mu}{\delta}{\nu}}
C_{{\varepsilon}\ \, {\sigma}\ \,;{\beta}}^{\ \, {\mu}\ \, {\nu}} 
u^{\varepsilon}u^{\sigma}
\nonumber \\ &&
- \frac{3}{20}
C_{{\beta}{\gamma}{\mu}{\delta};{\nu}} C_{{\varepsilon}\ \, {\sigma}}^{\ \,
  {\mu}\ \, {\nu}} 
u^{\varepsilon}u^{\sigma}
+ \frac{1}{3}
\left[
  \frac{1}{2} C_{{\mu}{\nu}{\gamma}{\lambda}} C^{{\mu}{\nu}\ \,{\lambda}}_{\ \ \
    {\delta}\ \,\, ;{\beta}} + 
C_{{\mu}{\varepsilon}{\gamma}{\lambda}} C^{{\mu}\ \ \, {\lambda}}_{\ \,
{\sigma}{\delta}\ \,\, ;{\beta}} u^{\varepsilon}u^{\sigma}\right] 
\nonumber \\ &&
- \frac{19}{60}\left[
  \frac{1}{2} C_{{\mu}{\nu}{\gamma}{\lambda}} C^{{\mu}{\nu}\ \ \, \,;{\lambda}}_{\ \ \
    {\delta}{\beta}} + 
C_{{\mu}{\varepsilon}{\gamma}{\lambda}} C^{{\mu}\ \ \ \ ;{\lambda}}_{\ \,
{\sigma}{\delta}{\beta}} u^{\varepsilon}u^{\sigma}\right]
\Bigg\} {\Delta}{\tau}^3 + O({\Delta}{\tau}^4).
\label{eq:finalans}
\end{eqnarray}

\section{Some special cases}
\label{s:cases}
Our expression (\ref{eq:finalans}) for the quasi-local contribution to the
self-force is quite general, applying to a massive particle with any
4-velocity in any vacuum background spacetime. There are, however, cases which
are of more intrinsic interest than others. In particular, cases in which the
background is a black hole are of interest for developing templates for LISA.
We now examine the form that $f_{\rm QL}^\alpha$ takes for several particle
four-velocities in a Schwarzschild background and for two
simple four-velocities in a Kerr background.

We begin with the Schwarzschild background. In standard Schwarzschild
coordinates, the line element is
\begin{equation}
   ds^2=-\left(1-\frac{2m}{r}\right)dt^2+\frac{dr^2}{1-\frac{2m}{r}}+r^2
   d\theta^2 + r^2 \sin^2\theta d\phi^2, 
   \label{eq:SchwazMet}
\end{equation}
where $m$ is the mass of the Schwarzschild black hole.  Consider a general
particle four velocity. Because of the spherical symmetry of the background,
we can always arrange that at any instant, the particle as well as the spatial
projection of the particle's 4-velocity be in the equatorial plane. Doing so,
we have that $\theta=\pi/2$ and $u^\theta=0$ for a general particle
4-velocity. This leave three non-vanishing components for $u^\alpha$, however,
the normalization condition $u^\alpha u_\alpha = -1$ can be used to eliminate
one (we have chosen to eliminate $u^t$). Thus, in this background, the
velocity of of the particle is completely specified in general by a choice of
$u^r=dr/d\tau=\dot{r}$ and $u^\phi=d\phi/d\tau=\dot{\phi}$.

Putting this general 4-velocity and the Schwarzschild metric into
(\ref{eq:finalans}) yields
\begin{align}
   f_{\rm QL}^t=\;&\mu^2 \frac{m^2}{r^7(r-2m)}\,\sqrt{r\left(r\,\dot{r}^2
         +\left(r-2m\right)\left({\dot{\phi}}^2\,r^2+1\right)\right)}\\
      &\times\biggl[9\,r^3\,{\dot{\phi}}^2\left(2\,{\dot{\phi}}^2\,r^2+1
         \right)\Delta\tau^2+\frac{3}{10}\,\dot{r}\left(64\,{\dot{\phi}}^2
         \,r^2+150\,{\dot{\phi}}^4\,r^4+1\right)\Delta\tau^3 + 
         O(\Delta\tau^4)\biggr]\\
   f_{\rm QL}^r=\;&\mu^2\frac{m}{r^8}\,\biggl[
   \begin{aligned}[t]
      &-9\,r^4\left(2\,{\dot{\phi}}^2\,r^2+1\right){\dot{\phi}}^2\,\dot{r}
         \,\Delta\tau^2\\
      &+\frac{3}{20}\,\biggl(
      \begin{aligned}[t]
         &62\,{\dot{\phi}}^2\,r^2m+80\,m\,{\dot{\phi}}^4\,r^4+4\,m
            -40\,{\dot{\phi}}^4\,r^5-2\,r\,{\dot{r}}^2\\
         &-300\,r^5\,{\dot{r}}^2\,{\dot{\phi}}^4 -2\,r-128\,r^3\,{\dot{r}}^2
            \,{\dot{\phi}}^2-31\,{\dot{\phi}}^2\,r^3\biggr)\,\Delta\tau^3 
            + O(\Delta \tau^4)\biggr]
      \end{aligned}
   \end{aligned}\\
   f_{\rm QL}^\theta=\;&0\\
   f_{\rm QL}^\phi=\;&-\mu^2\frac{m^2}{r^7}\,\dot{\phi}\,\biggl[9\,r\,
      \left({\dot{\phi}}^2\,r^2+1\right)\left(2\,{\dot{\phi}}^2\,r^2+1\right)
      \Delta\tau^2+\frac{3}{20}\,\left(388\,{\dot{\phi}}^2\,r^2+300\,
      {\dot{\phi}}^4\,r^4+99\right)\dot{r}\,\Delta\tau^3+O(\Delta\tau^4)\biggr]
\end{align}
One expects, of course, the vanishing of the $\theta$ component by symmetry
arguments. However, we notice several other features. For purely radial motion
($\dot{\phi}=0$), the quasi-local part of the self-force vanishes to $O(\Delta
\tau^3)$, and the $f_{\rm QL}^\phi$ component vanishes to $O(\Delta \tau^4)$ -
indeed, by symmetry arguments it must vanish to all orders.  Thus, the
quasi-local part of the self-force is directed toward the black hole to order
$O(\Delta \tau^4)$ in this case.  For purely tangential motion ($\dot{r}=0$),
$f_{\rm QL}^r$ vanishes to $O(\Delta \tau^2)$. Further, there is no $\Delta
\tau^3$ contribution to $f_{\rm QL}^t$ or $f_{\rm QL}^\phi$ in this case, On
the other hand, $f_{\rm QL}^r$ does not vanish even if for a static particle,
where $\dot{\phi}=\dot{r}=0$. 

A particularly interesting case is that of a particle following a circular
geodesic around the black hole. In this case, 
\begin{equation}
   \dot{r}=0, ~ ~ ~ ~ ~ ~ ~~ ~~ ~~ ~~ ~     
   \dot{\phi}=\frac{1}{r}\sqrt{\frac{m}{r-3m}}, 
\end{equation}
which gives a quasi-local self-force contribution of
\begin{align}
   f_{\rm QL}^t=\;&9\,\mu^2\,\frac{m^3}{r^6}\,\sqrt{\frac{r}{r-3\,m}}\,
      \frac{\left(r-m\right)}{\left(r-3\,m\right)^2}\,\Delta\tau^2 
      + O(\Delta \tau^4)\\
   f_{\rm QL}^r=\;&\frac{3}{20}\,\mu^2\,\frac{m^2}{r^8}\,\left(35\,m^2-19\,mr
      -2\,r^2 \right)\frac{\left(r-2\,m \right)}{\left(r-3\,m \right)^2}\,
      \Delta \tau^3 +O(\Delta \tau^4)\\
   f_{\rm QL}^\theta=\;&0\\
   f_{\rm QL}^\phi=\;&-9\,\mu^2\,\frac{m^2}{r^7}\,\left(r-2\,m\right) 
      \sqrt{\frac{m}{r-3\,m}}\,\frac{\left(r-m\right)}{\left(r-3\,m \right)^2}
      \, \Delta \tau^2 + O(\Delta \tau^4)
\end{align}
Interestingly, we can assign physical meanings to two of these components.
Assuming that the self-force causes an adiabatic deviation from the background
geodesic, we can define an energy for the particle of $E=u_t$ and an angular
momentum of $L=u_\phi$. Thus, $dE/d\tau=du^t/d\tau=f^t/\mu$ and
$dL/d\tau=dy^\phi/d\tau=f^\phi/\mu$. In other words, $f_{\rm QL}^t/\mu$ and
$f_{\rm QL}^\phi$ can respectively be interpreted as the energy and angular
momentum radiated by the particle due to the quasi-local part of the
self-force.

Of course, we expect astrophysical black holes to provide a Kerr background
spacetime, so this background is more astrophysically relevant than
Schwarzschild. Unfortunately, as is often the case, calculations in the Kerr
background lead to longer and less managable expressions. Using symbolic
algebra programs, it is straightforward to calculate, for example, for the
case of general equatorial motion in Kerr, and we have done so using MAPLE.
The expressions, however, are so unwieldy that we choose not to display them
here. Rather, we illustrate with the two simplest motions in a Kerr
background, stationary with respect to an observer at rest at infinity and
co-rotating with the black hole. We restrict the particle to be in the
equatorial plane in both cases for convenience.

We start with the metric in Boyer-Lindquist coordinates, 
\begin{equation}
   ds^2=-\left( \frac{\Delta - a^2\sin^2\theta}{\Sigma} \right) dt^2 
   - 2a\sin^2\theta\left( \frac{r^2+a^2-\Delta}{\Sigma} \right) dt d\phi
   +\left[ \frac{\left( r^2+a^2 \right)^2-a^2\Delta\sin^2\theta}{\Sigma}
   \right]\sin^2\theta d\phi^2 + \frac{\Sigma}{\Delta} dr^2 + \Sigma
   d\theta^2,
\end{equation}
where
\begin{align}
   \Sigma=\,&r^2+a^2\cos^2\theta,\\
   \Delta=\,&r^2+a^2-2mr.
\end{align}
As usual, $a$ is the spin parameter for the Kerr black hole and $m$ is its
mass. We will only be considering the special case where the particle is
located at $\theta=\pi/2$, and only particles with $u^r=u^\theta=0$. Again, we
have the normalization condition $u^\alpha u_\alpha = -1$ which we use to
further eliminate $u^t$. We therefore have only one specifiable 4-velocity
component, $\dot{\phi}$.

We begin with the case where $\dot{\phi}=0$. This is the case of a particle which
is momentarily stationary with respect to an observer at spatial infinity. In
that case, the quasi-local part of the self-force becomes
\begin{align}
   f_{\rm QL}^t=\,&18\,\mu^2\,\frac{ a^2m^3}{r^{10}}\,
      \frac{\left(r^2-2\,mr+2\,a^2\right)}{\left(r^2-2\,mr+a^2\right)^3}\,
      \frac{\left(r \left(r-2\,m\right) \left(r^2-2\,mr+a^2\right)^2 
      \right)^{3/2} }{ \left(r-2\,m\right)^4 }\,\Delta\tau^2
      +O(\Delta\tau^4),\\
   f_{\rm QL}^r=\,&-\frac{3}{10}\,\mu^2\,\frac{m^2}{r^{11}} \,
      \frac{\left(r^2-2\,mr+a^2\right)}{\left( r-2\,m \right)^2} 
      \left( 10\,a^4-16\,mra^2+8\,a^2r^2-4\,r^3m+r^4+4\,m^2r^2 \right) 
      \,\Delta\tau^3+O(\Delta\tau^4) ,\\
   f_{\rm}^\theta=\,&0,\\
   f_{\rm QL}^\phi=\,&-9\,\mu^2\, \frac{am^2}{r^{10} }\frac{ \left(
      r^2-2\,mr+2\,a^2 \right)}{ \left(r^2-2\,mr+a^2\right)^3 }\,
      \frac{\left( r \left(r-2\,m\right)  \left(r^2-2\,mr+a^2\right)^2 
      \right)^{3/2}}{ \left(r-2\,m\right)^3 }\,\Delta\tau^2+O(\Delta \tau^4) .
\end{align}
It is interesting to note that there is a self-force at order $\Delta \tau^2$
in this case, unlike the case of a static particle in Schwarzschild. 

One might argue that this is not unexpected, since in Kerr the particle is
``moving'' with respect to the rotating background. A fairer comparison,
therefore, might be with a particle that is co-rotating in the Kerr
background, i.e. one for which $u_\phi=0$. However, in this case we find the
quasi-local part of the self-force is
\begin{align}
   f_{\rm QL}^r=\,&-\frac{3}{10}\,\mu^2\,\frac{m^2}{r^{11}} 
      \frac{ \left( r^2-2\,mr+a^2 \right)}{\left( r^3+2\,ma^2+a^2r \right)^2}
      \begin{aligned}[t]
         \biggl( & 22\,a^4m^2r^2-27\,a^2r^5m-71\,a^4r^3m-44\,a^6rm\\
         &+r^8+10\,r^6a^2+27\,a^4r^4+28\,a^6r^2+10\,a^8 \biggr)\,
            \Delta\tau^3+O(\Delta \tau^4) 
      \end{aligned}
   \\
   f_{\rm QL}^\phi=\,&-9\,\mu^2\,\frac{m^2 a}{r^9}
      \begin{aligned}[t]
         &\frac{ \left( r^2+a^2 \right) \left( r^4+3\,a^2r^2-2\,a^2mr +2\,a^4
         \right)}{\left(r^2-2\,mr+a^2\right)\left(r^3+a^2r+2\,ma^2\right)^4}\\
         &\times\,\left( r \left( r^3+a^2r+2\,ma^2 \right) 
            \left( r^2-2\,mr+a^2 \right)  \right) ^{3/2} \,\Delta\tau^2 
            + O(\Delta \tau^4) 
      \end{aligned}
\end{align}
Thus, in both analogues of the static particle in a Schwarzschild background,
the rotation of the Kerr background induces a radiation reaction force at
order $\Delta \tau^3$.

We end this section with a warning. While the expressions derived here might
be useful for comparisons, they do not, in general, have any intrinsic meaning
(the case of the circular geodesic in Schwarzschild being somewhat of an
exception). One obvious reason for this is that we have found only a part of
the self-force which reflects a very limited part of the particle's world-line.
However, there is a more subtle limitation as well. The self-force is a gauge
dependent quantity, depending on the perturbation gauge chosen. Our
expressions, which are composed of quantities that are gauge invariants of the
background, are nonetheless tied to the Lorentz perturbation gauge. Thus,
these expressions may, in general, only be legitimately compared to other expressions for the self-force in the Lorentz gauge.
\section{Conclusion}
\label{sec:conc}
In this paper, we have discussed a novel approach to calculating the
self-force experienced by a massive particle moving on a geodesic in a curved
background. In this approach, proposed by Poisson and Wiseman, one does not
regularize the retarded Green's functions nor the retarded field. Rather, one
explicitly calculates the tail part of the retarded Green's function in the
normal neighborhood of its current position. From this, one can calculate the
``quasi-local'' part of the self-force, which arises from some finite portion
of the particle's worldline within this neighborhood, of duration $\Delta
\tau$.  The rest of the self-force can then be obtained using the full
retarded Green's function (without need of regularization), since it is well
behaved (and gives the correct contribution) for the portion of the particle's
world-line beyond $\Delta \tau$ to the past.

We have also carried out the first step in this procedure, the calculation of
the first two non-vanishing terms of an expansion of the quasi-local part of
the self-force in $\Delta \tau$. Our expression has some remarkable
properties. First, it is quite general, in that it does not rely on prior
specification of the particle motion nor on prior specification of the
background geometry (we have restricted our attention to vacuum backgrounds in
this paper, but this was for convenience, and exactly the same procedure can
be used for calculating the quasi-local part of the self-force in backgrounds
with matter). Second, we express the quasi-local part of the self-force in
terms of quantities that are purely local to the particle. Thus, our
expression does not require a detailed understanding of the past history of
the particle - it is only a function of the current position and velocity of
the particle. This can hardly be surprising - within a normal neighborhood,
there is a unique geodesic specified by any 4-vector at a given point, which
the particle is assumed to be travelling along.

Much work remains in order to determine even if the Poisson-Wiseman
prescription is feasible, let alone to calculate a complete self-force using
it. While the approach here is quite general, it is likely that the retarded
Green's function from which the rest of the self-force is calculated will be
most easily obtained in most scenarios of interest by a mode-sum expansion.
It remains to be seen whether it is technically feasible to calculate the
quasi-local part of the self-force to sufficient order that it has sufficient
precision at a distance from the particle at which the mode-sum converges
well. Work is currently under way to begin addressing such questions\cite{AW}.

Some hope arises from the work of Anderson and Hu\cite{A&H04}, who have
shown how to calculate the tail part of the retarded Green's function for a
scalar particle in Schwarzschild using the Hadamard-WKB approximation. This
approach, which should work for spin-1 (electromagnetic) and spin-2
(gravitational) fields on a Schwarzschild background, allows one to calculate
to a much higher order in $\Delta \tau$ for a fixed amount of effort.
Furthermore, it might be extendible to Kerr backgrounds. In any case, we echo
their sentiment, that this approach warrants further investigation.

\section*{Acknowledgments}
This work was supported in part by the Center for Gravitational Wave
Physics, which is funded by the National Science Foundation under
Cooperative Agreement PHY 0114375.  EEF was supported in part by NSF
grants PHY-9722189 and PHY-0140209, by the Sloan Foundation and by the
Radcliffe Institute. WGA was supported by the Center for Gravitational Wave
Astronomy, which is funded by the National Aeronautic and Space Administration
through the NASA University Research Center Program, and through NSF grants
PHY-0140326 and PHY-0200852. We would like to thank the Capra Gang, and in
particular Eric Poisson, Alan Wiseman, and Leor Barack, for useful 
conversations.

\appendix

\section{Geometric content of the expansion coefficients}
\label{sec:Counting}

In this appendix, we use symmetry and dimensional arguments to deduce
the forms of the coefficients 
$f_{\alpha}^{(0)}$, $f_{\alpha}^{(1)}$, $f_{\alpha}^{(2)}$ and $f_{\alpha}^{(3)}$
appearing in the 
expansion (\ref{eq:forceexpand}) of the quasi-local piece of the self
force, up to 
some unknown numerical coefficients.
We start by noting that the self force is proportional to the
square ${\mu}^2$ of the particle's mass, from Eq.\ (\ref{eq:QLint}).  
Factoring out this factor of ${\mu}^2$, we can write the expansion
(\ref{eq:forceexpand}) as 
\begin{equation}
   f_{QL\,{\alpha}}({\tau},{\Delta}{\tau})
   ={\mu}^2 \left[ {\hat f}_{\alpha}^{(0)} +{\hat f}_{\alpha}^{(1)} {\Delta}
   {\tau}+{\hat f}_{\alpha}^{(2)} {\Delta}{\tau}^2+{\hat f}_{\alpha}^{(3)} {\Delta}
   {\tau}^3+{\calO}({\Delta}{\tau}^4) \right].  
\label{eq:forceexpand1}
\end{equation}
Here the coefficients ${\hat f}_{\alpha}^{(j)} = f_{\alpha}^{(j)}/{\mu}^2$ satisfy
the following key properties:
\begin{itemize}
\item They are independent of the particle mass ${\mu}$.

\item They must be constructed as polynomial expressions in the
following tensors at $x^{\alpha}({\tau})$: the metric $g_{{\alpha}{\beta}}$, the
four velocity $u^{\alpha}$, and the Weyl tensor and its various
symmetrized derivatives $C_{{\alpha}{\beta}{\gamma}{\delta}}$,
$C_{{\alpha}{\beta}{\gamma}{\delta};{\varepsilon}}$,
$C_{{\alpha}{\beta}{\gamma}{\delta};({\varepsilon}{\rho})}$, etc.  The Ricci tensor
does not appear since we are assuming a vacuum background, and the
four acceleration of the curve $a^{\alpha}$ does not appear since we are
assuming a geodesic curve at zeroth order.  The Levi-Civita tensor
$\lvepsilon_{{\alpha}{\beta}{\gamma}{\delta}}$ cannot appear, since the
expression (\ref{eq:QLint}) for the self force is invariant under the parity
transformation $\lvepsilon_{{\alpha}{\beta}{\gamma}{\delta}} \to -
\lvepsilon_{{\alpha}{\beta}{\gamma}{\delta}}$.  

\item Since the force is dimensionless (in geometric units in which $G
= c = 1$), and ${\Delta}{\tau}$ has dimensions of length, each
coefficient ${\hat f}_{\alpha}^{(j)}$ has dimension $({\rm
length})^{-2-j}$.   

\item Each coefficient ${\hat f}_{\alpha}^{(j)}$ must be orthogonal to the
four velocity $u^{\alpha}$, since the total force (\ref{eq:QLint}) has this
property. 

\end{itemize}

We now apply these properties to deduce the most general allowed forms
of the various coefficients, which are given in Eqs.\ (\ref{eq:f0ans}) --
(\ref{eq:f3ans}) above. 

\subsection{The zeroth order coefficient ${\hat f}_{\alpha}^{(0)}$}
\label{sec:0order}

The coefficient ${\hat f}_{\alpha}^{(0)}$ has dimension $({\rm
length})^{-2}$, and hence must 
be linear in $C_{{\alpha}{\beta}{\gamma}{\delta}}$ which also has dimension
$({\rm length})^{-2}$.  None of the derivatives of the Weyl tensor can
appear.  However, there is no non-vanishing vector that can be
constructed out of $C_{{\alpha}{\beta}{\gamma}{\delta}}$, $g_{{\alpha}{\beta}}$
and $u^{\alpha}$ that is orthogonal to $u^{\alpha}$.  One needs to
contract at least three of the indices on the Weyl tensor with the
metric or the four velocity.  However the 
Weyl tensor is traceless on all pairs of indices, so one cannot use
the metric to contract any pair of indices.  Also the antisymmetry
properties $C_{({\alpha}{\beta}){\gamma}{\delta}} =
C_{{\alpha}{\beta}({\gamma}{\delta})} =0$ of the Weyl tensor mean that one
cannot contract with three factors of four velocity.  Hence 
the coefficient ${\hat f}_{\alpha}^{(0)}$ must vanish, {\it cf.}\ 
Eq.\ (\ref{eq:f0ans}) above.
A version of this argument was first given by Ori, and
was used to deduce the fact that the formula for the gravitational
self force could contain only the tail term and could not contain any
local terms, unlike the scalar and electromagnetic self-force
expressions \cite{Ori95}.

An alternative, simpler version of the argument can be obtained by
considering the independent, electric and magnetic components of the
Weyl tensor.  Introduce an orthonormal basis $e^{\alpha}_{{\hat 0}} =
u^{\alpha}$ and $e^{\alpha}_{{\hat j}}$, $1 \le j \le 3$.  Then we
define
\begin{equation}
{\cal E}_{{\hat i}{\hat j}} = C_{{\hat 0}{\hat i}{\hat 0}{\hat j}}
{\equiv}e_{\hat 0}^{\alpha}e_{\hat i}^{\beta}e_{\hat 0}^{\gamma}e_{\hat
  j}^{\delta}C_{{\alpha}{\beta}{\gamma}{\delta}}
\label{eq:calEdef}
\end{equation}
and 
\begin{equation}
{\cal B}_{{\hat i}{\hat j}} = - \frac{1}{2} \lvepsilon_{{\hat i}{\hat
    k}{\hat l}} C_{{\hat k}{\hat l}{\hat 0}{\hat j}} {\equiv}
- \frac{1}{2} \lvepsilon_{{\alpha}{\beta}{\gamma}{\delta}} C^{{\gamma}{\delta}}_{\ \
    \ {\varepsilon}{\rho}} u^{\alpha}e_{\hat i}^{\beta}u^{\varepsilon}e_{\hat j}^{\rho}.
\label{eq:calBdef}
\end{equation}
In vacuum these are symmetric, traceless tensors, and the Weyl tensor
can be expressed in terms of these tensors via the formulae
\begin{equation}
C_{{\hat 0}{\hat i}{\hat j}{\hat k}} = - \lvepsilon_{{\hat l}{\hat
    j}{\hat k}} {\cal B}_{{\hat l}{\hat i}}
\label{eq:calBdef1}
\end{equation}
and
\begin{equation}
C_{{\hat i}{\hat j}{\hat k}{\hat l}} = - \lvepsilon_{{\hat i}{\hat
    j}{\hat p}} \lvepsilon_{{\hat k}{\hat l}{\hat q}} {\cal E}_{{\hat
    p}{\hat q}}
\label{eq:calEdef1}
\end{equation}
together with Eq.\ (\ref{eq:calEdef}).
The coefficient ${\hat f}_{\hat i}^{(0)}$ must be a three-vector that is
linear in ${\cal E}_{{\hat i}{\hat j}}$, or linear in ${\cal B}_{{\hat i}{\hat
j}}$ with one factor of $\lvepsilon_{{\hat i}{\hat j}{\hat k}}$ by
parity arguments, and there is no such vector since ${\cal B}_{[{\hat
i}{\hat j}]}=0$.

\subsection{The first order coefficient ${\hat f}_{\alpha}^{(1)}$}
\label{sec:1order}

The coefficient ${\hat f}_{\alpha}^{(1)}$ has dimension $({\rm
length})^{-3}$ and so must be linear in
$C_{{\alpha}{\beta}{\gamma}{\delta};{\varepsilon}}$.  Again, it is impossible to
form a non-vanishing 4-vector that is orthogonal to $u^{\alpha}$ by
contracting $C_{{\alpha}{\beta}{\gamma}{\delta};{\varepsilon}}$ with the metric
and/or four velocity.  The additional derivative index ${\varepsilon}$
cannot be contracted with any of the indices on the Weyl tensor, since
by the Bianchi identity the Weyl tensor is divergence free on all its
indices in vacuum: 
\begin{equation}
C_{{\alpha}{\beta}{\gamma}{\delta}}^{\ \, \ \ \ \ ;{\delta}} =0.
\label{eq:identity89}
\end{equation}
The ${\varepsilon}$ index
can be contracted with the four velocity to form
$C_{{\alpha}{\beta}{\gamma}{\delta};{\varepsilon}} u^{\varepsilon}$, but then one is
faced with the same problem as above of contracting three of the four
remaining free indices to obtain a vector.  It follows that 
${\hat f}_{\alpha}^{(1)}$ must vanish, {\it cf.}\ Eq.\ (\ref{eq:f1ans})
above.

One can also phrase this argument in terms of the electric and magnetic
components of the Weyl tensor as before.  The components of
$C_{{\alpha}{\beta}{\gamma}{\delta};{\varepsilon}}$ can be represented as the
``time derivatives'' and ``spatial derivatives'' of ${\cal E}_{{\hat
 i}{\hat j}}$ and ${\cal   B}_{{\hat i}{\hat j}}$:
\begin{eqnarray}
\label{eq:dotcalEdef}
{\dot {\cal E}}_{{\hat i}{\hat j}} &{\equiv}&
C_{{\alpha}{\beta}{\gamma}{\delta};{\varepsilon}} u^{\alpha}e_{\hat i}^{\beta}
u^{\gamma}e_{\hat j}^{\delta}u^{\varepsilon}\\
{\cal E}_{{\hat i}{\hat j},{\hat k}} &{\equiv}&
C_{{\alpha}{\beta}{\gamma}{\delta};{\varepsilon}} u^{\alpha}e_{\hat i}^{\beta}
u^{\gamma}e_{\hat j}^{\delta}e_{\hat k}^{\varepsilon}
\label{eq:gradcalEdef}
\\
{\dot {\cal B}}_{{\hat i}{\hat j}} &{\equiv}&
- \frac{1}{2} \lvepsilon_{{\alpha}{\beta}{\gamma}{\delta}} C^{{\gamma}{\delta}}_{\ \
    \ {\varepsilon}{\rho};{\sigma}} u^{\alpha}e_{\hat i}^{\beta}u^{\varepsilon}
e_{\hat j}^{\rho}u^{\sigma}. 
\\
{\cal B}_{{\hat i}{\hat j},{\hat k}} &{\equiv}&
- \frac{1}{2} \lvepsilon_{{\alpha}{\beta}{\gamma}{\delta}} C^{{\gamma}{\delta}}_{\ \
    \ {\varepsilon}{\rho};{\sigma}} u^{\alpha}e_{\hat i}^{\beta}u^{\varepsilon}
e_{\hat j}^{\rho}e_{\hat k}^{\sigma}. 
\label{eq:gradcalBdef}
\end{eqnarray}
These quantities are not all independent but obey constraints that
follow from the Bianchi identity:
\begin{eqnarray}
\label{eq:calEdiv}
{\cal E}_{{\hat i}{\hat j},{\hat j}} &=&0,
\\
{\cal B}_{{\hat i}{\hat j},{\hat j}} &=&0,
\\
{\dot {\cal E}}_{{\hat i}{\hat j}} &=& -{\epsilon}_{{\hat i}{\hat
    k}{\hat l}} \, {\cal B}_{{\hat j}{\hat k},{\hat l}},
\label{eq:dotcalEformula}
\\
{\dot {\cal B}}_{{\hat i}{\hat j}} &=& {\epsilon}_{{\hat i}{\hat
    k}{\hat l}} \, {\cal E}_{{\hat j}{\hat k},{\hat l}}.
\label{eq:dotcalBformula}
\end{eqnarray}
Since the time derivatives can be obtained from the spatial
derivatives, we can without loss of generality restrict attention to
the spatial derivatives.  Thus, 
the coefficient ${\hat f}_{\hat i}^{(1)}$ must depend linearly on
${\cal E}_{{\hat i}{\hat j},{\hat k}}$ 
and/or ${\cal B}_{{\hat i}{\hat j},{\hat k}}$, and can depend in
addition only on 
${\delta}_{{\hat i}{\hat j}}$ and on $\lvepsilon_{{\hat i}{\hat j}{\hat
k}}$.  Consider first the derivative ${\cal E}_{{\hat i}{\hat j},{\hat k}}$.  
This quantity cannot be contracted with $\lvepsilon_{{\hat i}{\hat
j}{\hat k}}$ by parity arguments, and it is easy to see that one
cannot obtain a nonvanishing spatial vector.  
The only candidate vectors are the divergence
$
{\cal E}_{{\hat i}{\hat j},{\hat k}} {\delta}^{{\hat j}{\hat k}}
$
which vanishes by Eq.\ (\ref{eq:calEdiv}), and the contraction
$
{\cal E}_{{\hat i}{\hat j},{\hat k}} {\delta}^{{\hat i}{\hat j}},
$
which vanishes by the traceless property of ${\cal E}_{{\hat i}{\hat
j}}$.  Next, consider the derivative ${\cal B}_{{\hat i}{\hat j},{\hat
k}}$.   By parity, this quantity must be 
accompanied by one factor of
$\lvepsilon_{{\hat i}{\hat j}{\hat k}}$ (or an odd number of factors
of it).  Any such tensor will have an even number of indices, and
consequently it is impossible to obtain by contraction a vector.

\subsection{The second order coefficient ${\hat f}_{\alpha}^{(2)}$}
\label{sec:2order}

The coefficient ${\hat f}_{\alpha}^{(2)}$ has dimension $({\rm
length})^{-4}$, and so must be either quadratic in 
$C_{{\alpha}{\beta}{\gamma}{\delta}}$ or linear in
$C_{{\alpha}{\beta}{\gamma}{\delta};({\varepsilon}{\sigma})}$.
Consider first the second derivative term
$C_{{\alpha}{\beta}{\gamma}{\delta};({\varepsilon}{\sigma})}$.  
Let us first analyze the term
$C_{{\alpha}{\beta}{\gamma}{\delta};{\varepsilon}{\sigma}}$ without the
symmetrization.  The six indices on this tensor must be reduced to one
index by contractions with the metric and/or the four velocity.  As
before, the problem is getting rid of at least three of the indices
${\alpha}$, ${\beta}$, ${\gamma}$ and ${\delta}$.  The only new type of
contraction that is available is to contract one of these indices with
the second derivative index ${\sigma}$ [contractions with the first
derivative index ${\varepsilon}$ vanish by Eq.\ (\ref{eq:identity89})].
However, such a contraction can be re-expressed as a product of two
Weyl tensors by commuting the indices.  The same argument also applies
to the symmetrized derivative
$C_{{\alpha}{\beta}{\gamma}{\delta};({\varepsilon}{\sigma})}$, since it is a linear
combination of un-symmetrized derivatives.
Thus, it is sufficient to
consider expressions that are quadratic in the Weyl tensor \footnote{
In fact, the only nontrivial candidate expression one can construct from the
symmetrized second derivative vanishes.  This expression 
is
$$
C_{{\alpha}{\beta}{\gamma}{\delta};({\varepsilon}{\sigma})} g^{{\sigma}{\delta}} u^{\beta}
u^{\gamma}u^{\varepsilon}= - \frac{1}{2} \left[ 
C_{{\varepsilon}{\sigma}{\alpha}}^{\ \ \ \ {\rho}}
C_{{\rho}{\beta}{\gamma}}^{\ \ \ \ {\sigma}}
+C_{{\varepsilon}{\sigma}{\beta}}^{\ \ \ \ {\rho}}
C_{{\alpha}{\rho}{\gamma}}^{\ \ \ \ {\sigma}}
+C_{{\varepsilon}{\sigma}{\gamma}{\rho}}
C_{{\alpha}{\beta}}^{\ \ \ {\rho}{\sigma}}
\right]
g^{{\sigma}{\delta}} u^{\beta}
u^{\gamma}u^{\varepsilon},
$$
where we have used Eq.\ (\protect{\ref{eq:identity89}}).  Here the last term in
the square brackets vanishes since the the first factor is symmetric
in the index pair $({\rho}{\sigma})$ by virtue of being contracted with
$u^{\varepsilon}u^{\gamma}$, and the second factor is antisymmetric in
$({\rho}{\sigma})$.  The remaining two terms in the square brackets cancel
against each other.}.  

Consider therefore expressions that are quadratic in the Weyl tensor, or
linear in $C_{{\alpha}{\beta}{\gamma}{\delta}}
C_{{\varepsilon}{\sigma}{\rho}{\lambda}}$.  We can classify such expressions
in terms of the number of contractions between indices on the first
Weyl tensor and indices on the second Weyl tensor.  The cases of zero,
four and one contractions are easy to dispense with.  For example, in the
case of one contraction,
there is no vector orthogonal to $u^{\alpha}$ that can be obtained by
contracting 
$C_{{\alpha}{\beta}{\gamma}{\delta}} C_{{\varepsilon}{\sigma}{\rho}}^{\ \ \ \, 
{\delta}}$ with $g_{{\alpha}{\beta}}$ and $u^{\alpha}$ (without further
contractions between the two Weyl tensors).  
In the case of three contractions, there are two
different tensors that one can construct,
namely
\begin{equation}
V_{{\alpha}{\beta}} = C_{{\alpha}{\gamma}{\delta}{\varepsilon}} C_{\beta}^{\
  \,{\gamma}{\delta}{\varepsilon}} 
\end{equation}
and
\begin{equation}
W_{{\alpha}{\beta}} = C_{{\alpha}{\gamma}{\delta}{\varepsilon}} C_{\beta}^{\
  \,{\delta}{\gamma}{\varepsilon}}.
\end{equation}
However it follows from $C_{{\alpha}[{\beta}{\gamma}{\delta}]}=0$ that
$V_{{\alpha}{\beta}} = 2 W_{{\alpha}{\beta}}$, so there is only one
independent tensor obtainable from three contractions.  Furthermore
using the formulae 
(\ref{eq:calEdef}), (\ref{eq:calBdef1}) and (\ref{eq:calEdef1}) one
can show that $W_{{\alpha}{\beta}}$ is proportional to the metric:
\begin{equation}
W_{{\alpha}{\beta}} = \frac{1}{4} W_{\gamma}^{\ \, {\gamma}} g_{{\alpha}{\beta}}.
\end{equation}
Hence there is no non-vanishing vector that can be formed using three
contractions.  

Consider next the case of two contractions.  There are three different
four index tensors that can be obtained with two contractions, namely
\begin{eqnarray}
X_{{\alpha}{\beta}{\varepsilon}{\sigma}} &=& C_{{\alpha}{\beta}{\gamma}{\delta}}
C_{{\varepsilon}{\sigma}}^{\ \  \, {\gamma}{\delta}}, \\
Y_{{\alpha}{\gamma}{\varepsilon}{\rho}} &=& C_{{\alpha}{\beta}{\gamma}{\delta}}
C_{{\varepsilon}\ \,{\rho}}^{\ \, {\beta}\ \, {\delta}},
\end{eqnarray}
and
\begin{equation}
Z_{{\alpha}{\gamma}{\varepsilon}{\sigma}} = C_{{\alpha}{\beta}{\gamma}{\delta}}
C_{{\varepsilon}{\sigma}}^{\ \ \ {\beta}{\delta}}.
\end{equation}
The tensor $X_{{\alpha}{\beta}{\varepsilon}{\sigma}}$ does not yield any
candidate expressions, since one needs to contract it with
three factors of the four velocity, and it is antisymmetric on the index pairs
$({\alpha}{\beta})$ and $({\varepsilon}{\sigma})$.  
Similarly the tensor $Z_{{\alpha}{\gamma}{\varepsilon}{\sigma}}$ is
antisymmetric on the index pairs $({\alpha}{\gamma})$ and
$({\varepsilon}{\sigma})$, and so does not yield any candidate expressions.
The tensor
$Y_{{\alpha}{\gamma}{\varepsilon}{\rho}}$ can be used to construct
the non-vanishing 4-vector $Y_{{\alpha}{\gamma}{\varepsilon}{\rho}} u^{\gamma}
u^{\varepsilon}u^{\rho}$.  Projecting this vector orthogonal to the four
velocity yields the possible term
\begin{equation}
({\delta}_{\alpha}^{\beta}+ u_{\alpha}u^{\beta})
C_{{\beta}{\gamma}{\delta}{\varepsilon}} C_{{\sigma}\ \, {\rho}}^{\ \, {\gamma}\, \
  {\varepsilon}} u^{\delta}u^{\sigma}u^{\rho}.
\end{equation}
This quantity is nonvanishing in general and and so can appear in the
expression for ${\hat f}_{\alpha}^{(2)}$, {\it cf.}\ Eq.\
(\ref{eq:f2ans}) above.  It is the only term that
arises at this order in ${\Delta}{\tau}$.  Using the formulae (\ref{eq:calEdef})
and (\ref{eq:calBdef1}), we can express this term in terms of the
electric and magnetic components of the Weyl tensor as
\begin{equation}
- {\epsilon}_{{\hat i}{\hat j}{\hat k}} {\cal E}_{{\hat j}{\hat l}}
  {\cal B}_{{\hat l}{\hat k}}.
\label{eq:ansc}
\end{equation}
This completes the derivation of the most general allowed form of
${\hat f}_{\alpha}^{(2)}$.   

The analysis of expressions that are quadratic in the Weyl tensor can
be rephrased more simply in terms of the tensors
${\cal E}_{{\hat i}{\hat j}}$ and ${\cal B}_{{\hat i}{\hat j}}$.  
One needs a spatial vector that is bilinear in ${\cal E}_{{\hat i}{\hat
j}}$ and/or ${\cal B}_{{\hat i}{\hat j}} {\epsilon}_{{\hat p}{\hat
q}{\hat r}}$, since by parity each factor of ${\cal B}_{{\hat i}{\hat
j}}$ must be accompanied by a factor of ${\epsilon}_{{\hat p}{\hat q}{\hat r}}$.
It is easy to see that the only  
non-vanishing candidate expression is the product (\ref{eq:ansc}).

\subsection{The third order coefficient ${\hat f}_{\alpha}^{(3)}$}
\label{sec:3order}

The coefficient ${\hat f}_{\alpha}^{(3)}$ has dimension $({\rm
length})^{-5}$, and so must be either bilinear in 
$C_{{\alpha}{\beta}{\gamma}{\delta}}$ and
$C_{{\alpha}{\beta}{\gamma}{\delta};{\varepsilon}}$, or else 
or linear in the symmetrized third derivative
$C_{{\alpha}{\beta}{\gamma}{\delta};({\varepsilon}{\sigma}{\rho})}$.
The argument used in the first paragraph of Sec.\ \ref{sec:2order}
above also applies here 
and shows that any expression constructed from the third derivative
can be expressed as a product of the Weyl tensor and a first
derivative of the Weyl tensor.  Therefore it is sufficient to consider
such products.

For analyzing these products, a fully covariant analysis would be very
complex, so we 
use the simpler formalism of the electric and magnetic components.
We need to construct a vector that is a contraction of a product of one of the
tensors
\begin{equation}
{\cal E}_{{\hat i}{\hat j}},\ \ \ \ {\cal B}_{{\hat i}{\hat j}}
{\epsilon}_{{\hat p}{\hat q}{\hat r}}
\label{eq:t1}
\end{equation}
together with one of the tensors
\begin{equation}
\label{eq:t2}
{\cal E}_{{\hat i}{\hat j},{\hat k}},\ \ \ \ 
{\cal B}_{{\hat i}{\hat j},{\hat k}} {\epsilon}_{{\hat p}{\hat q}{\hat
    r}}.
\end{equation}
As before we can neglect the time derivatives because of Eqs.\
(\ref{eq:dotcalEformula}) and (\ref{eq:dotcalBformula}).  
It follows from the identities (\ref{eq:calEdiv}) --
(\ref{eq:dotcalBformula}) and the fact that 
all of the tensors (\ref{eq:t1}) and (\ref{eq:t2})
are symmetric and tracefree on the index pair
$({\hat i}{\hat j})$
that there are only four possible nonvanishing vectors that
one can construct, namely\footnote{One can also construct from the
  time derivatives the possible terms 
${\epsilon}_{{\hat i}{\hat j}{\hat k}} {\dot {\cal B}}_{{\hat j}{\hat l}}
	{\cal E}_{{\hat l}{\hat k}}$
and
${\epsilon}_{{\hat i}{\hat j}{\hat k}} {\cal B}_{{\hat j}{\hat l}} 
{\dot {\cal E}}_{{\hat l}{\hat k}}$.  However these terms can be
expressed in terms of the four quantities
(\protect{\ref{eq:candidates}}) using the formulae
(\protect{\ref{eq:dotcalEformula}}) and
(\protect{\ref{eq:dotcalBformula}}):
$$
{\epsilon}_{{\hat i}{\hat j}{\hat k}} {\dot {\cal B}}_{{\hat j}{\hat l}}
	{\cal E}_{{\hat l}{\hat k}} = 
{\cal E}_{{\hat j}{\hat k}} {\cal E}_{{\hat j}{\hat k},{\hat i}} - 
{\cal E}_{{\hat j}{\hat k}} {\cal E}_{{\hat i}{\hat j},{\hat k}},
\ \ \ \ \ \ \ \ \ 
{\epsilon}_{{\hat i}{\hat j}{\hat k}} {\cal B}_{{\hat j}{\hat l}} 
{\dot {\cal E}}_{{\hat l}{\hat k}} = 
{\cal B}_{{\hat j}{\hat k}} {\cal B}_{{\hat j}{\hat k},{\hat i}} -  
{\cal B}_{{\hat j}{\hat k}} {\cal B}_{{\hat i}{\hat j},{\hat k}}. 
$$
}
\begin{equation}
\label{eq:candidates}
{\cal E}_{{\hat j}{\hat k}} {\cal E}_{{\hat j}{\hat k},{\hat i}},\ \ \
\ 
{\cal E}_{{\hat j}{\hat k}} {\cal E}_{{\hat i}{\hat j},{\hat k}},\ \ \
\ 
{\cal B}_{{\hat j}{\hat k}} {\cal B}_{{\hat j}{\hat k},{\hat i}},\ \ \
\ 
{\rm and} \ \ \ \ 
{\cal B}_{{\hat j}{\hat k}} {\cal B}_{{\hat i}{\hat j},{\hat k}}. 
\end{equation}
The corresponding covariant expressions can be obtained from Eqs.\
(\ref{eq:calEdef}) and (\ref{eq:calBdef}) and are, respectively,
\begin{eqnarray}
({\delta}_{\alpha}^{\beta}+ u_{\alpha}u^{\beta}) C_{{\gamma}{\mu}{\delta}{\nu}}
C_{{\varepsilon}\ \, {\sigma}\ \,;{\beta}}^{\ \, {\mu}\ \, {\nu}} u^{\gamma}u^{\delta}
u^{\varepsilon}u^{\sigma},\\
C_{{\alpha}{\gamma}{\mu}{\delta};{\nu}} C_{{\varepsilon}\ \, {\sigma}}^{\ \, {\mu}\ \, {\nu}}
u^{\gamma}u^{\delta}u^{\varepsilon}u^{\sigma},\\
({\delta}_{\alpha}^{\beta}+ u_{\alpha}u^{\beta}) u^{\gamma}u^{\delta}\left[
  \frac{1}{2} C_{{\mu}{\nu}{\gamma}{\lambda}} C^{{\mu}{\nu}\ \,{\lambda}}_{\ \ \
    {\delta}\ \,\, ;{\beta}} + 
C_{{\mu}{\varepsilon}{\gamma}{\lambda}} C^{{\mu}\ \ \, {\lambda}}_{\ \,
{\sigma}{\delta}\ \,\, ;{\beta}} u^{\varepsilon}u^{\sigma}\right], \\
{\rm and} \ \ \ \ \ \ 
u^{\gamma}u^{\delta}\left[
  \frac{1}{2} C_{{\mu}{\nu}{\gamma}{\lambda}} C^{{\mu}{\nu}\ \ \, \,;{\lambda}}_{\ \ \
    {\delta}{\alpha}} + 
C_{{\mu}{\varepsilon}{\gamma}{\lambda}} C^{{\mu}\ \ \ \ ;{\lambda}}_{\ \,
{\sigma}{\delta}{\alpha}} u^{\varepsilon}u^{\sigma}\right].
\end{eqnarray}
{\it cf.}\ Eq.\ (\ref{eq:f3ans}) above.

\section{Local expansions of some fundamental bitensors}
\label{sec:foundations}

In this appendix we compute local covariant expansions of
a number of fundamental bitensors, including the second derivative
${\sigma}_{;{\alpha}{\beta}}$ of the world function, various covariant
derivatives of the parallel displacement bivector $g_{\alpha}^{\
\,{\alpha}'}$, and the Van-Vleck Morette determinant ${\Delta}$, to order
$O(s^5)$ beyond the leading order.  These expansions were originally
computed to $O(s^4)$ by Christensen \cite{C76,C78}, and extended to
$O(s^5)$ by Brown and Ottewill \cite{BO83,BO86}.  Our results agree
with those of Brown and Ottewill, except for one case where we correct
their result [Eq.\ (\ref{eq:boxdelta0}) below].
All of the formulae in this appendix are valid in any number of
spacetime dimensions, except in specific cases, noted below, which
make use of the identity (\ref{eq:identity4}) which is specific to four 
dimensions.  For a detailed and pedagogic review of the computational
methods used here see Ref.\ \cite{Poisson03}.

We first note a number of useful identities that follow from the
Bianchi identity and from properties of the Riemann tensor:
\begin{eqnarray}
\label{eq:identity1}
R_{{\alpha}{\beta}{\gamma}{\delta}} R_{\varepsilon}^{\ \, {\delta}{\gamma}{\beta}} &=& \frac{1}
{2} R_{{\alpha}{\beta}{\gamma}{\delta}} R_{\varepsilon}^{\ \,
{\beta}{\gamma}{\delta}} \\
\label{eq:identity2}
R_{{\alpha}{\beta}{\gamma}{\delta}} R_{{\varepsilon}\ \ \ \ \, ;{\mu}}^{\ \,
{\delta}{\gamma}{\beta}} &=& \frac{1}{2}
R_{{\alpha}{\beta}{\gamma}{\delta}} R_{{\varepsilon}\ \ \ \ \, ;{\mu}}^{\ \,
{\beta}{\gamma}{\delta}} \\
\label{eq:identity3}
R_{\alpha}^{\ \ {\mu}{\nu}{\lambda}} R_{{\beta}{\mu}{\gamma}{\nu};{\lambda}} &=& -
\frac{1}{2} R_{\alpha}^{\ \ {\mu}{\nu}{\lambda}}
R_{{\beta}{\mu}{\nu}{\lambda};{\gamma}} \\
\label{eq:identity4}
C_{{\alpha}{\mu}{\nu}{\lambda}} C_{\beta}^{\ \, {\mu}{\nu}{\lambda}} &=& \frac{1}{4}
g_{{\alpha}{\beta}} \, C_{{\mu}{\nu}{\lambda}{\rho}} C^{{\mu}{\nu}{\lambda}{\rho}}.
\end{eqnarray}
Next, we derive the coincidence limits $x^{\prime}\to x$ of symmetrized
derivatives of the world function ${\sigma}$.
Following Ref.\ \cite{DB60}, these can be obtained by repeatedly
differentiating the identity   
\begin{equation}
{\sigma}= \frac{1}{2} g^{{\alpha}{\beta}} {\sigma}_{;{\alpha}} {\sigma}_{;{\beta}}
\label{eq:identity00}
\end{equation}
and by taking the coincidence limit.  The results are
\begin{equation}
\lim_{x^{\prime}\to x} {\sigma}_{;{\alpha}{\beta}} = g_{{\alpha}{\beta}},
\label{eq:coinc1}
\end{equation}
\begin{equation}
\lim_{x^{\prime}\to x} {\sigma}_{;{\alpha}{\beta}{\gamma}} = 0,
\end{equation}
\begin{equation}
   \lim_{x^{\prime}\to x} {\sigma}_{;{\alpha}{\beta}{\gamma}{\delta}} = - \frac{1}{3} \left[ R_{{\alpha}{\gamma}{\beta}{\delta}} + R_{{\alpha}{\delta}{\beta}{\gamma}} \right], 
\label{eq:coinc3}
\end{equation}
\begin{equation}
\lim_{x^{\prime}\to x} {\sigma}_{;{\alpha}{\beta}{\gamma}{\delta}{\varepsilon}} = - \frac{1}
{4} \left[ R_{{\alpha}{\gamma}{\beta}{\delta};{\varepsilon}}
+ R_{{\alpha}{\delta}{\beta}{\gamma};{\varepsilon}} +
R_{{\alpha}{\delta}{\beta}{\varepsilon};{\gamma}} +
R_{{\alpha}{\varepsilon}{\beta}{\delta};{\gamma}} +
R_{{\alpha}{\varepsilon}{\beta}{\gamma};{\delta}} +
R_{{\alpha}{\gamma}{\beta}{\varepsilon};{\delta}} \right], 
\label{eq:coinc4}
\end{equation}
\begin{equation}
\lim_{x^{\prime}\to x} {\sigma}_{;{\alpha}{\beta}({\gamma}{\delta}{\varepsilon}{\mu})} = -
\frac{12}{5} R_{{\alpha}{\gamma}{\beta}{\delta};{\varepsilon}{\mu}} - \frac{8}{15}
R_{{\alpha}{\gamma}{\delta}{\chi}} R_{{\beta}{\varepsilon}{\mu}}^{\ \ \ \ {\chi}}, 
\label{eq:coinc5}
\end{equation}
and
\begin{equation}
\lim_{x^{\prime}\to x} {\sigma}_{;{\alpha}{\beta}({\gamma}{\delta}{\varepsilon}{\mu}{\nu})}
= - \frac{10}{3} R_{{\alpha}{\gamma}{\beta}{\delta};{\varepsilon}{\mu}{\nu}} -
\frac{10}{3} R_{{\alpha}{\gamma}{\delta}{\chi}} R_{{\beta}{\varepsilon}{\mu}\ \
;{\nu}}^{\ \ \ \ 
{\chi}}.
\label{eq:coinc6}
\end{equation}
In Eqs.\ (\ref{eq:coinc5}) -- (\ref{eq:coinc6}), the right hand sides
are to be symmetrized over the index pair 
$({\alpha}{\beta})$ and over as many of the indices
$({\gamma}{\delta}{\varepsilon}{\mu}{\nu})$ as are present. 

Next, consider any smooth bitensor $T(x,x^{\prime})$ (we suppress tensor indices
on $T$).  We can expand this bitensor about $x$ as a covariant Taylor
expansion 
\begin{equation}
\label{eq:taylor}
T(x,x^{\prime}) = \sum_{n=0}^{\infty}\frac{1}{n!}
t_{n}^{{{\alpha}_1}\ldots{{\alpha}_n}}(x) {\sigma}_{;{\alpha}_1}(x,x^{\prime}) \ldots
{\sigma}_{;{\alpha}_n}(x,x^{\prime}),
\end{equation}
where the coefficients $t_n^{{{\alpha}_1}\ldots{{\alpha}_n}}$ are local
tensors at $x$.  By repeatedly differentiating Eq.\ (\ref{eq:taylor})
and taking the coincidence limit we find for $n=0$ that 
\begin{equation}
t_{0}(x) = \lim_{x^{\prime}\to x} T(x,x^{\prime}),
\end{equation}
together with the recursion relation for $n\ge1$
\begin{equation}
t_{n\,{{\alpha}_1}\ldots{{\alpha}_n}}(x) = \lim_{x^{\prime}\to x}
T_{;({{\alpha}_1}\ldots{{\alpha}_n})}(x,x^{\prime}) - \sum_{r=0}^{n-1} 
\left(\begin{array}{c} n \\ r \end{array}\right)
t_{r\,({{\alpha}_1}\ldots{{\alpha}_r};{{\alpha}_{r+1}}\ldots{{\alpha}_{n}})}(x).
\label{eq:recursion}
\end{equation}
We now apply the recursion relation (\ref{eq:recursion}) to the
bitensor ${\sigma}_{;{\alpha}{\beta}}(x,x^{\prime})$, using the formulae
(\ref{eq:coinc1}) -- (\ref{eq:coinc6}).  The result is the expansion 
\begin{eqnarray}
{\sigma}_{;{\alpha}{\beta}} &=& g_{{\alpha}{\beta}} 
- \frac{1}{3} R_{{\alpha}{\gamma}{\beta}{\delta}} {\sigma}^{;{\gamma}}
{\sigma}^{;{\delta}}
+ \frac{1}{12} R_{{\alpha}{\gamma}{\beta}{\delta};{\varepsilon}} {\sigma}^{;{\gamma}}
{\sigma}^{;{\delta}} {\sigma}^{;{\varepsilon}}
- \left[ \frac{1}{60} R_{{\alpha}{\gamma}{\beta}{\delta};{\varepsilon}{\mu}} +
\frac{1}{45} R_{{\alpha}{\gamma}{\delta}{\rho}} R_{{\beta}{\varepsilon}{\mu}}^{\ \ \
\ {\rho}} \right] {\sigma}^{;{\gamma}}
{\sigma}^{;{\delta}} {\sigma}^{;{\varepsilon}} {\sigma}^{;{\mu}} \nonumber \\
\mbox{} && + 
\left[ \frac{1}{360} R_{{\alpha}{\gamma}{\beta}{\delta};{\varepsilon}{\mu}{\nu}} +
\frac{1}{120} R_{{\alpha}{\gamma}{\delta}{\rho};{\nu}} R_{{\beta}{\varepsilon}{\mu}}^{\ \ \
\ {\rho}} 
+ \frac{1}{120} R_{{\alpha}{\gamma}{\delta}{\rho}} R_{{\beta}{\varepsilon}{\mu}\ \
;{\nu}}^{\ \ \ \ {\rho}} \right] {\sigma}^{;{\gamma}}
{\sigma}^{;{\delta}} {\sigma}^{;{\varepsilon}} {\sigma}^{;{\mu}} {\sigma}^{;{\nu}}
+ {\calO}(s^6).
\label{eq:sigmaexpand}
\end{eqnarray}
It follows from Eq. (\ref{eq:sigmaexpand}) that in vacuum
\begin{equation}
{\Box}{\sigma}= n + {\calO}(s^4),
\label{eq:expand3}
\end{equation}
where $n$ is the number of spacetime dimensions, and that
\begin{equation}
   {\Box}({\sigma}^{;{\alpha}})={\calO}(s^3).
\label{eq:expand6}
\end{equation}

A similar computation can be carried out for the 
tensor $Q_{{\alpha}{\beta}{\gamma}}$ defined by
\begin{equation}
Q_{{\alpha}{\beta}{\gamma}} {\equiv}g_{\alpha}^{\ \, {\alpha}^{\prime}} \,
g_{{{\alpha}^{\prime}}{\beta};{\gamma}}.
\label{eq:Qdef}
\end{equation}
By starting with the identity \cite{DB60}
\begin{equation}
g^{{\beta}{\gamma}} g_{{\alpha}^{\prime}{\alpha};{\beta}} {\sigma}_{;{\gamma}} =0,
\label{eq:identity0}
\end{equation}
repeatedly differentiating and taking the coincidence limit one obtains
\begin{equation}
\lim_{x^{\prime}\to x} g_{{\alpha}^{\prime}{\alpha};{\beta}} = 0,
\label{eq:coinc1a}
\end{equation}
\begin{equation}
\lim_{x^{\prime}\to x} 
g_{{\alpha}^{\prime}{\alpha};{\beta}{\gamma}} 
\, g^{{\alpha}^{\prime}}_{\ \ {\rho}}
= -\frac{1}{2} R_{{\alpha}{\rho}{\beta}{\gamma}},
\end{equation}
\begin{equation}
\lim_{x^{\prime}\to x} 
g_{{\alpha}^{\prime}{\alpha};{\beta}({\gamma}{\delta})} 
\, g^{{\alpha}^{\prime}}_{\ \ {\rho}}
= -\frac{2}{3} R_{{\alpha}{\rho}{\beta}{\gamma};{\delta}},
\end{equation}
\begin{equation}
\lim_{x^{\prime}\to x} 
g_{{\alpha}^{\prime}{\alpha};{\beta}({\gamma}{\delta}{\varepsilon})} 
\, g^{{\alpha}^{\prime}}_{\ \ {\rho}}
= -\frac{3}{4} R_{{\alpha}{\rho}{\beta}{\gamma};{\delta}{\varepsilon}}
- \frac{1}{4} R_{{\alpha}{\rho}{\gamma}{\varphi}} R_{{\beta}{\delta}{\varepsilon}}^{\ \
\ \ {\varphi}},
\end{equation}
and
\begin{equation}
\lim_{x^{\prime}\to x} 
g_{{\alpha}^{\prime}{\alpha};{\beta}({\gamma}{\delta}{\varepsilon}{\mu})} 
\, g^{{\alpha}^{\prime}}_{\ \ {\rho}}
= -\frac{4}{5} R_{{\alpha}{\rho}{\beta}{\gamma};{\delta}{\varepsilon}{\mu}}
- \frac{8}{15} R_{{\alpha}{\rho}{\gamma}{\varphi};{\mu}} R_{{\beta}{\delta}{\varepsilon}}^{\ \
\ \ {\varphi}}
- \frac{3}{5} R_{{\alpha}{\rho}{\gamma}{\varphi}} R_{{\beta}{\delta}{\varepsilon}\ \ ;{\mu}}^{\ \
\ \ {\varphi}}.
\label{eq:coinc5a}
\end{equation}
In Eqs.\ (\ref{eq:coinc1a}) -- (\ref{eq:coinc5a}), the right hand
sides are understood to be symmetrized over as many of the indices
$({\gamma}{\delta}{\varepsilon}{\mu})$ as are present.
Combining the coincidence limits (\ref{eq:coinc1a}) --
(\ref{eq:coinc5a}) with the formulae (\ref{eq:taylor}) --
(\ref{eq:recursion}) for Taylor expansions 
together with the definition (\ref{eq:Qdef}) one obtains
\begin{eqnarray}
Q_{{\alpha}{\beta}{\gamma}}
&=&
\frac{1}{2} R_{{\alpha}{\beta}{\gamma}{\delta}} {\sigma}^{;{\delta}}
- \frac{1}{6} R_{{\alpha}{\beta}{\gamma}{\delta};{\varepsilon}} 
{\sigma}^{;{\delta}} {\sigma}^{;{\varepsilon}}  
+ \frac{1}{24} \left[ R_{{\alpha}{\beta}{\gamma}{\delta};{\varepsilon}{\mu}} +
R_{{\alpha}{\beta}{\delta}{\rho}} R_{{\gamma}{\varepsilon}{\mu}}^{\ \ \ \ {\rho}} \right]  
{\sigma}^{;{\delta}} {\sigma}^{;{\varepsilon}} {\sigma}^{;{\mu}} \nonumber \\
\mbox{} && - \left[ \frac{1}{120}
R_{{\alpha}{\beta}{\gamma}{\delta};{\varepsilon}{\mu}{\nu}} 
+ \frac{7}{360} R_{{\alpha}{\beta}{\delta}{\rho};{\nu}} R_{{\gamma}{\varepsilon}{\mu}}^{\
\ \ \ {\rho}} + \frac{1}{60} R_{{\alpha}{\beta}{\delta}{\rho}}
R_{{\gamma}{\varepsilon}{\mu}\ \, ;{\nu}}^{\ \ \ \ {\rho}} \right]
{\sigma}^{;{\delta}} {\sigma}^{;{\varepsilon}} {\sigma}^{;{\mu}} {\sigma}^{;{\nu}}
+ {\calO}(s^5).
\label{eq:Qexpand}
\end{eqnarray}

Next, we derive an expansion for the quantity $g_{\alpha}^{\ \, 
{{\alpha}^{\prime}}} \, g_{{{\alpha}^{\prime}}{\beta};{\gamma}}^{\ \ \ \ \ \
;{\gamma}}$ by using the identity \cite{DB60}
\begin{equation}
g_{\alpha}^{\ \, {{\alpha}^{\prime}}} \,
g_{{{\alpha}^{\prime}}{\beta};{\gamma}}^{\ \ \ \ \ \ ;{\gamma}} =
Q_{{\alpha}{\beta}{\gamma}}^{\ \ \ \ \ ;{\gamma}} - Q^{\delta}_{\ \,
{\alpha}{\gamma}} Q_{{\delta}{\beta}}^{\ \ \ {\gamma}},
\end{equation} 
together with the expansion
(\ref{eq:Qexpand}).  The result is
\begin{eqnarray}  
g_{\alpha}^{\ \, {{\alpha}^{\prime}}} \,
g_{{{\alpha}^{\prime}}{\beta};{\gamma}}^{\ \ \ \ \ \ ;{\gamma}} &=&
- \frac{1}{4} C_{{\alpha}{\rho}{\gamma}{\varphi}} 
C_{{\beta}\ {\delta}}^{\ \,{\rho}\ {\varphi}} \, {\sigma}^{;{\gamma}} {\sigma}^{;{\delta}}
\nonumber \\
\mbox{} && + \left[ 
\frac{1}{20}
C_{{\alpha}{\rho}{\gamma}{\varphi}} 
C_{{\beta}\ {\delta}\ \ ;{\varepsilon}}^{\ \,{\rho}\ {\varphi}} 
+ \frac{7}{60}
C_{{\alpha}{\rho}{\gamma}{\varphi};{\varepsilon}} 
C_{{\beta}\ {\delta}}^{\ \,{\rho}\ {\varphi}} 
+ \frac{2}{45}
C_{{\alpha}{\beta}{\rho}{\gamma};{\varphi}} 
C_{{\delta}\ {\varepsilon}}^{\ \,{\rho}\ {\varphi}}  \right]
{\sigma}^{;{\gamma}} {\sigma}^{;{\delta}} {\sigma}^{;{\varepsilon}}  
+ {\calO}(s^4),
\label{eq:Zexpand}
\end{eqnarray}
where we have specialized to the vacuum case $R_{{\alpha}{\beta}}=0$.

Next, following Ref.\ \cite{DB60} we define the tensor 
\begin{equation}
D_{{\alpha}{\beta}} = - g_{\alpha}^{\ {{\alpha}^{\prime}}}
{\sigma}_{;{{\alpha}^{\prime}}{\beta}}
\end{equation}
which is related to the Van Vleck-Morette determinant \cite{vanvleck}
${\Delta}$ by  
\begin{equation}
{\Delta}= {\rm det} D_{\alpha}^{\ \, {\beta}}.
\label{eq:vanvleckdef}
\end{equation}
Using the identity \cite{DB60}
\begin{equation}
D_{{\alpha}{\beta}} = Q_{{\alpha}{\gamma}{\beta}} \, {\sigma}^{;{\gamma}} +
{\sigma}_{;{\alpha}{\beta}}
\end{equation}
together with the expansions (\ref{eq:sigmaexpand}) and
(\ref{eq:Qexpand}) we obtain the expansion
\begin{eqnarray}
D_{{\alpha}{\beta}} &=& g_{{\alpha}{\beta}} 
+ \frac{1}{6} R_{{\alpha}{\gamma}{\beta}{\delta}} {\sigma}^{;{\gamma}}
{\sigma}^{;{\delta}}
- \frac{1}{12} R_{{\alpha}{\gamma}{\beta}{\delta};{\varepsilon}} {\sigma}^{;{\gamma}}
{\sigma}^{;{\delta}} {\sigma}^{;{\varepsilon}}
+ \left[ \frac{1}{40} R_{{\alpha}{\gamma}{\beta}{\delta};{\varepsilon}{\mu}} +
\frac{7}{360} R_{{\alpha}{\gamma}{\delta}{\rho}} R_{{\beta}{\varepsilon}{\mu}}^{\ \ \
\ {\rho}} \right] {\sigma}^{;{\gamma}}
{\sigma}^{;{\delta}} {\sigma}^{;{\varepsilon}} {\sigma}^{;{\mu}} \nonumber \\
\mbox{} && - 
\left[ \frac{1}{180} R_{{\alpha}{\gamma}{\beta}{\delta};{\varepsilon}{\mu}{\nu}} +
\frac{1}{90} R_{{\alpha}{\gamma}{\delta}{\rho};{\nu}} R_{{\beta}{\varepsilon}{\mu}}^{\ \ \
\ {\rho}} 
+ \frac{1}{120} R_{{\alpha}{\gamma}{\delta}{\rho}} R_{{\beta}{\varepsilon}{\mu}\ \
;{\nu}}^{\ \ \ \ {\rho}} \right] {\sigma}^{;{\gamma}}
{\sigma}^{;{\delta}} {\sigma}^{;{\varepsilon}} {\sigma}^{;{\mu}} {\sigma}^{;{\nu}}
+ {\calO}(s^6).
\label{eq:Dexpand}
\end{eqnarray}
The determinant of $D_{\alpha}^{\ \ {\beta}}$ can be calculated via (\ref{e:DetTr}).
Taking the square root of the determinant gives
\begin{eqnarray}
{\Delta}^{1/2}  &=& 1 + \frac{1}{12} R_{{\alpha}{\beta}} {\sigma}^{;{\alpha}}
{\sigma}^{;{\beta}} 
- \frac{1}{24} R_{{\alpha}{\beta};{\gamma}} {\sigma}^{;{\alpha}}
{\sigma}^{;{\beta}} {\sigma}^{;{\gamma}}
+ \left[ \frac{1}{80} R_{{\alpha}{\beta};{\gamma}{\delta}} + \frac{1}{360}
R_{{\rho}{\alpha}{\beta}{\varphi}} R^{{\rho}\ \ \ {\varphi}}_{\ \, {\gamma}{\delta}} 
+ \frac{1}{288} R_{{\alpha}{\beta}} R_{{\gamma}{\delta}} 
\right] {\sigma}^{;{\alpha}} {\sigma}^{;{\beta}} {\sigma}^{;{\gamma}}
{\sigma}^{;{\delta}}
\nonumber \\
\mbox{} && - \left[ \frac{1}{360} R_{{\alpha}{\beta};{\gamma}{\delta}{\varepsilon}}
+ \frac{1}{288} R_{{\alpha}{\beta}} R_{{\gamma}{\delta};{\varepsilon}} 
+ \frac{1}{360} 
R_{{\rho}{\alpha}{\beta}{\varphi}} R^{{\rho}\ \ \ {\varphi}}_{\ \, {\gamma}{\delta}\
\, ;{\varepsilon}} 
\right]  {\sigma}^{;{\alpha}} {\sigma}^{;{\beta}} {\sigma}^{;{\gamma}}
{\sigma}^{;{\delta}} {\sigma}^{;{\varepsilon}} + {\calO}(s^6).
\label{eq:Deltaexpand}
\end{eqnarray}
Acting on this expression with the wave operator and using the identities
(\ref{eq:identity1}) -- (\ref{eq:identity3}) together with Eq.\
(\ref{eq:sigmaexpand}) we obtain
\begin{eqnarray}
{\Box}{\Delta}^{1/2} &=& \frac{1}{6} R + \bigg[ \frac{1}{40} {\Box}
R_{{\alpha}{\beta}} -\frac{1}{120} R_{;{\alpha}{\beta}} + \frac{1}{72}
R \, R_{{\alpha}{\beta}} - \frac{1}{30} R_{{\alpha}{\gamma}} R^{\gamma}_{\ \,
{\beta}} 
+ \frac{1}{60} R_{{\alpha}{\gamma}{\beta}{\delta}} R^{{\gamma}{\delta}}
+ \frac{1}{60} R_{{\alpha}{\gamma}{\delta}{\varepsilon}} R_{\beta}^{\ \, 
{\gamma}{\delta}{\varepsilon}} \bigg] \, {\sigma}^{;{\alpha}} {\sigma}^{;{\beta}} 
\nonumber \\
\mbox{} && + \bigg[ \frac{1}{45} R_{{\alpha}{\mu}} R_{{\beta}\ \ ;{\gamma}}^{\ \ {\mu}} 
- \frac{1}{180} R_{{\mu}{\nu}} R_{{\alpha}\ \, {\beta}\ \, ;{\gamma}}^{\ \, {\mu}\ \, {\nu}}
-\frac{1}{120} R_{{\alpha}{\beta}\ \ ;{\mu}{\gamma}}^{\ \ \ ;{\mu}}
-\frac{1}{180} R_{{\mu}{\nu};{\alpha}} R_{{\beta}\ \, {\gamma}}^{\ \, {\mu}\ \, {\nu}} 
\nonumber \\ \mbox{} &&
- \frac{1}{90} R_{{\alpha}{\mu}{\nu}{\lambda}} R_{{\beta}\ \ \ \ \ ;{\gamma}}^{\ \
{\mu}{\nu}{\lambda}} 
- \frac{1}{144} R R_{{\alpha}{\beta};{\gamma}}
+ \frac{1}{360} R_{;{\alpha}{\beta}{\gamma}}
\bigg] \, {\sigma}^{;{\alpha}} {\sigma}^{;{\beta}} {\sigma}^{;{\gamma}}
+ {\calO}(s^4). 
\label{eq:boxdelta0}
\end{eqnarray}
Specializing to the vacuum case $R_{{\alpha}{\beta}}=0$ yields
\begin{equation}
{\Box}{\Delta}^{1/2}= \frac{1}{60} C_{{\alpha}{\mu}{\nu}{\lambda}}
C_{\beta}^{\ \ {\mu}{\nu}{\lambda}} \, {\sigma}^{;{\alpha}}
{\sigma}^{;{\beta}} - 
 \frac{1}{90}  C_{{\alpha}{\mu}{\nu}{\lambda}} C_{{\beta}\ \ \ \ \, ;{\gamma}}^{\ \,
{\mu}{\nu}{\lambda}}   \, {\sigma}^{;{\alpha}} {\sigma}^{;{\beta}}
{\sigma}^{;{\gamma}} + {\calO}(s^4). 
\label{eq:boxdelta}
\end{equation}
In the special case of 4 spacetime dimensions this can be further
simplified using the identity (\ref{eq:identity4}) to give
\begin{equation}
{\Box}{\Delta}^{1/2}= \frac{1}{240} C_{{\varepsilon}{\mu}{\nu}{\lambda}}
C^{{\varepsilon}{\mu}{\nu}{\lambda}} \, g_{{\alpha}{\beta}} \, {\sigma}^{;{\alpha}}
{\sigma}^{;{\beta}} - \frac{1}{360} C_{{\mu}{\nu}{\lambda}{\rho};{\gamma}}
C^{{\mu}{\nu}{\lambda}{\rho}} g_{{\alpha}{\beta}} 
{\sigma}^{;{\alpha}} {\sigma}^{;{\beta}} {\sigma}^{;{\gamma}} + {\calO}(s^4).
\label{eq:boxdelta1}
\end{equation}

\section{Covariant expansion of the retarded Green's function}
\label{sec:Vexpand}

In this appendix we derive the local covariant expansion of the
retarded Green's function $G_{\text{ret}}^{{\mu}{\nu}{{\alpha}'} {{\beta}'}}(x,x')$
which is defined by the differential equation
\begin{equation}
   ({\Box}g_{{\mu}{\alpha}}g_{{\nu}{\beta}}+2C_{{\mu}{\alpha}{\nu}{\beta}})
      G_{\text{ret}}^{{\mu}{\nu}{{\alpha}'} {{\beta}'}}(x,x')=-\left[
      g_{({\alpha}}{^{{\alpha}'}} g_{{\beta})}{^{{\beta}'}} + {\kappa}
      g_{{\alpha}{\beta}} g^{{\alpha}'{\beta}'} \right] {\delta}^4(x,x').
\label{eq:LGFE0}
\end{equation}
Here we have introduced a real parameter ${\kappa}$ to facilitate
comparison with the work of Allen, Folacci and Ottewill (AFO)
\cite{AFO}, who analyzed the case ${\kappa}= -1/2$.  For this paper we
are interested in the case ${\kappa}= 0$, {\it cf.} Eq.\
(\ref{eq:LGFE}) above.  Note that these two Green's functions, the
cases  ${\kappa}=0$ and ${\kappa}= -1/2$, are related to each other by
a trace reversal on the index pair $({\alpha}'{\beta}')$.
Throughout this appendix we specialize to 4 spacetime dimensions.

We use the standard method explained by Hadamard\cite{Hadamard23},
DeWitt and Brehme\cite{DB60}, and AFO \cite{AFO} in the scalar, vector
and tensor cases respectively, and we extend the expansions of AFO to
one higher order.  We assume for the Feynman Green's
function the expression
\begin{equation}
   G_{\rm F}^{{\mu}{\nu}{{\alpha}'} {{\beta}'}}(x,x')=
         \frac{1}{4{\pi}^2}
      \Bigl\{ \frac{U^{{\mu}{\nu}{{\alpha}'}{{\beta}'}}(x,x')}{{\sigma}+ i {\epsilon}}
         +V^{{\mu}{\nu}{{\alpha}'}{{\beta}'}}(x,x')~ \ln \left[ {\sigma}+ i
         {\epsilon}\right] + W^{{\mu}{\nu}{{\alpha}'}{{\beta}'}}(x,x')\Bigr\}
\label{eq:Feynman}
\end{equation}
for some bitensors $U^{{\mu}{\nu}{{\alpha}'}{{\beta}'}}$,
$V^{{\mu}{\nu}{{\alpha}'}{{\beta}'}}$ and $W^{{\mu}{\nu}{{\alpha}'}{{\beta}'}}$, where
we have introduced a 
regularization parameter ${\epsilon}$ to give the appropriate
singularity structure.  The expression (\ref{eq:Hadamard}) for the retarded
Green's function can be obtained by taking the negative of the
imaginary part of the Feynman Green's function (\ref{eq:Feynman}), and
by multiplying by the function ${\Theta}[\Sigma(x,x')]$ defined in
Sec.\ \ref{sec:quasilocal} \cite{DB60}.

Substituting the real part of the Green's function (\ref{eq:Feynman})
into the homogeneous version of the differential equation
(\ref{eq:LGFE0}), and equating to zero the coefficients of
$1/{\sigma}^2$, $\ln {\sigma}$, and the remainder gives the three
equations
\begin{equation}
U^{{\mu}{\nu}{{\alpha}'}{{\beta}'};{\gamma}} {\sigma}_{;{\gamma}} - {1 \over 2} 
U^{{\mu}{\nu}{{\alpha}'}{{\beta}'}} ( \ln {\Delta})^{;{\gamma}} {\sigma}_{;{\gamma}}
=0,
\label{eq:EE0}
\end{equation}
\begin{equation}
\mathfrak{D}_{{\alpha}{\beta}}{^{{\mu}{\nu}}}V_{{\mu}{\nu}{\alpha}'{\beta}'}
=0,
\label{eq:EE1}
\end{equation}
and
\begin{equation}
\mathfrak{D}_{{\alpha}{\beta}}{^{{\mu}{\nu}}}U_{{\mu}{\nu}{\alpha}'{\beta}'}  + 
2 V_{{\alpha}{\beta}{\alpha}'{\beta}'} 
+ 2 V_{{\alpha}{\beta}{\alpha}'{\beta}'}^{\ \ \ \ \ \ \ ;{\gamma}}
{\sigma}_{;{\gamma}} 
- V_{{\alpha}{\beta}{\alpha}'{\beta}'} (\ln {\Delta})^{;{\gamma}} {\sigma}_{;{\gamma}}
+ {\sigma}\, \mathfrak{D}_{{\alpha}{\beta}}{^{{\mu}{\nu}}}W_{{\mu}{\nu}{\alpha}'{\beta}'}
=0.
\label{eq:EE2}
\end{equation}
Here we have defined the differential operator 
\begin{equation}
   \mathfrak{D}_{{\alpha}{\beta}}{^{{\mu}{\nu}}}={\Box}{\delta}_{\alpha}^{\mu}
   {\delta}_{\beta}^{\nu}+ 2 C_{\alpha}{^{\mu}}{_{\beta}}{^{\nu}},
\label{eq:calDdef}
\end{equation}
and we have made use of the identity \cite{DB60}
\begin{equation}
{\Box}{\sigma}= 4 - (\ln {\Delta})^{;{\gamma}} {\sigma}_{;{\gamma}},
\label{eq:identity5}
\end{equation}
where ${\Delta}$ is the Van Vleck-Morette determinant (\ref{eq:vanvleckdef}).

The solution to the differential equation (\ref{eq:EE0}) for
$U^{{\alpha}{\beta}{{\alpha}'}{{\beta}'}}$ which 
is appropriate for the source term on the right hand side of Eq.\
(\ref{eq:LGFE0}) is
\begin{equation}
   U_{{\alpha}{\beta}{\alpha}'{\beta}'}={\Delta}^{1/2} \left[
      g_{{\alpha}'({\alpha}}g_{{\beta}){\beta}'}
      + {\kappa}\, g_{{\alpha}{\beta}}g_{{\alpha}'{\beta}'}\right],
\label{eq:U}
\end{equation}
where we have used the identity (\ref{eq:identity0}).

Next, we assume formal power series expansions for
$V_{{\alpha}{\beta}{{\alpha}'}{{\beta}'}}$ and
$W_{{\alpha}{\beta}{{\alpha}'}{{\beta}'}}$ of the form
\begin{equation}
   V_{{\alpha}{\beta}{\alpha}'{\beta}'}(x,x')=\sum_{n=0}^{\infty}
      V^n_{{\alpha}{\beta}{\alpha}'{\beta}'}(x,x') {\sigma}^n.
\label{eq:Vexpand}
\end{equation}
and
\begin{equation}
   W_{{\alpha}{\beta}{\alpha}'{\beta}'}(x,x')=\sum_{n=0}^{\infty}
      W^n_{{\alpha}{\beta}{\alpha}'{\beta}'}(x,x') {\sigma}^n.
\label{eq:Wexpand}
\end{equation}
Note that these expansions do not define unique representations of the
bitensors $V_{{\alpha}{\beta}{{\alpha}'}{{\beta}'}}$ and
$W_{{\alpha}{\beta}{{\alpha}'}{{\beta}'}}$, since the coefficients
$V^n_{{\alpha}{\beta}{{\alpha}'}{{\beta}'}}$ and 
$W^n_{{\alpha}{\beta}{{\alpha}'}{{\beta}'}}$ can be 
arbitrary functions of $x$ and $x^{\prime}$.  
However, one can obtain a unique set of coefficients
from the following prescription \cite{DB60}.  Pick a bisolution
$W^0_{{\alpha}{\beta}{{\alpha}'}{{\beta}'}}$ of the 
homogeneous wave equation
$\mathfrak{D}_{{\alpha}{\beta}}{^{{\mu}{\nu}}}W^0_{{\mu}{\nu}{\alpha}'{\beta}'}=0$.
Then, substitute the expansions (\ref{eq:Vexpand}) and
(\ref{eq:Wexpand}) into Eqs.\ (\ref{eq:EE1}) and (\ref{eq:EE2}),
simplify using the identities (\ref{eq:identity00}) and
(\ref{eq:identity5}), and equate to zero the 
coefficients of powers of ${\sigma}$.  The result is the following
recursive set of 
ordinary differential equations along the geodesic joining $x$ and
$x'$ that allow one to solve for the coefficients:
\begin{eqnarray}
\label{eq:recrel1}
V^0_{{\alpha}{\beta}{\alpha}'{\beta}'} + \left[
  V^0_{{\alpha}{\beta}{\alpha}'{\beta}';{\mu}} - {1 \over 2}
  V^0_{{\alpha}{\beta}{\alpha}'{\beta}'} (\ln {\Delta})_{;{\mu}} \right]
  {\sigma}^{;{\mu}} &=& - {1 \over 2}
  \mathfrak{D}_{{\alpha}{\beta}}{^{{\mu}{\nu}}} U_{{\mu}{\nu}{{\alpha}'}{{\beta}'}}, \\
V^n_{{\alpha}{\beta}{\alpha}'{\beta}'} + {1 \over n+1} \left[
  V^n_{{\alpha}{\beta}{\alpha}'{\beta}';{\mu}} - {1 \over 2}
  V^n_{{\alpha}{\beta}{\alpha}'{\beta}'} (\ln {\Delta})_{;{\mu}} \right]
  {\sigma}^{;{\mu}} &=& - {1 \over 2 n (n+1)}
  \mathfrak{D}_{{\alpha}{\beta}}{^{{\mu}{\nu}}} V^{n-1}_{{\mu}{\nu}{{\alpha}'}{{\beta}'}}
\label{eq:recrel2}
\end{eqnarray}
and
\begin{eqnarray}
W^n_{{\alpha}{\beta}{\alpha}'{\beta}'} + {1 \over n+1} \left[
  W^n_{{\alpha}{\beta}{\alpha}'{\beta}';{\mu}} - {1 \over 2}
  W^n_{{\alpha}{\beta}{\alpha}'{\beta}'} (\ln {\Delta})_{;{\mu}} \right]
  {\sigma}^{;{\mu}} &=& - {1 \over 2 n (n+1)}
  \mathfrak{D}_{{\alpha}{\beta}}{^{{\mu}{\nu}}}
  W^{n-1}_{{\mu}{\nu}{{\alpha}'}{{\beta}'}} \nonumber \\
&&- {1 \over n+1} V^{n}_{{\alpha}{\beta}{{\alpha}'}{{\beta}'}} + {1 \over 2 n^2
  (n+1) }  \mathfrak{D}_{{\alpha}{\beta}}{^{{\mu}{\nu}}}
  V^{n-1}_{{\mu}{\nu}{{\alpha}'}{{\beta}'}}.
\label{eq:recrel3}
\end{eqnarray}
Equations (\ref{eq:recrel2}) and (\ref{eq:recrel3}) apply for $n \ge 1$.
The power series (\ref{eq:Vexpand}) and (\ref{eq:Wexpand}) with these
coefficients converge in a neighborhood of the diagonal $x=x'$ for
analytic metrics \cite{Hadamard23}.

In this paper we are only interested in the coefficients
$V^n_{{\alpha}{\beta}{\alpha}'{\beta}'}$, which can be obtained from Eqs.\
(\ref{eq:recrel1}) and (\ref{eq:recrel2}). 
We expand each of these coefficients as
covariant Taylor series of the form (\ref{eq:taylor})
\begin{equation}
   V^n_{{\alpha}{\beta}{\alpha}'{\beta}'}=
g_{{\alpha}'}^{\ \ {\gamma}} g_{{\beta}'}^{\ \ {\delta}} \left[
      v^n_{{\alpha}{\beta}{\gamma}{\delta}}(x)
      +v^n_{{\alpha}{\beta}{\gamma}{\delta}{\varepsilon}}(x){\sigma}^{;{\varepsilon}}
      +\frac{1}{2}v^n_{{\alpha}{\beta}{\gamma}{\delta}{\varepsilon}{\zeta}}(x)
         {\sigma}^{;{\varepsilon}}{\sigma}^{;{\zeta}}
      +\frac{1}{6}v^n_{{\alpha}{\beta}{\gamma}{\delta}{\varepsilon}{\zeta}{\eta}}(x)
         {\sigma}^{;{\varepsilon}}{\sigma}^{;{\zeta}}{\sigma}^{;{\eta}}+\ldots\right],
\label{eq:CTSEVn}
\end{equation}
where the coefficients $v^n_{{\alpha}\ldots{\eta}}$ are local tensors at $x$.
We now specialize to the case $n=0$, and substitute the expansion
(\ref{eq:CTSEVn}) for $V^0_{{\alpha}{\beta}{\alpha}'{\beta}'}$ and the
expression (\ref{eq:U}) for $U_{{\alpha}{\beta}{\alpha}'{\beta}'}$ into the
differential equation (\ref{eq:recrel1}).  We simplify using the
identity (\ref{eq:identity0}) and the definition (\ref{eq:calDdef}),
and expand the various terms as power series in ${\sigma}_{;{\mu}}$ using
the expansions (\ref{eq:sigmaexpand}), (\ref{eq:Qexpand}),
(\ref{eq:Zexpand}), (\ref{eq:Deltaexpand}) and (\ref{eq:boxdelta1}).
Equating the coefficients of the various powers of ${\sigma}_{;{\mu}}$
then gives a series of equations that can be solved for the
coefficients $v^0_{{\alpha}\ldots{\eta}}$.  The results are
\begin{eqnarray}
   v^0_{{\alpha}{\beta}{\gamma}{\delta}}&=&-C_{{\alpha}{\gamma}{\beta}{\delta}},
\label{eq:v01}\\
   v^0_{{\alpha}{\beta}{\gamma}{\delta}{\varepsilon}}&=&\frac{1}{2}
      C_{{\alpha}{\gamma}{\beta}{\delta};{\varepsilon}},
\label{eq:v02}\\
   v^0_{{\alpha}{\beta}{\gamma}{\delta}{\varepsilon}{\zeta}}&=&
      -\frac{1}{3}C_{{\alpha}{\gamma}{\beta}{\delta};{\varepsilon}{\zeta}}
      -\frac{1}{6}C_{{\alpha}{\gamma}}{^{\mu}}{_{\varepsilon}}C_{{\beta}{\delta}{\mu}{\zeta}}
      +\frac{1}{6}g_{{\alpha}{\gamma}}C_{\beta}{^{{\mu}{\nu}}}{_{\varepsilon}}
         C_{{\delta}{\mu}{\nu}{\zeta}}
      -\frac{1}{180} {\Pi}_{{\alpha}{\beta}{\gamma}{\delta}} C^{{\mu}{\nu}{\rho}}{_{\varepsilon}}
         C_{{\mu}{\nu}{\rho}{\zeta}},
\label{eq:v03}
\end{eqnarray}
and
\begin{eqnarray}
   v^0_{{\alpha}{\beta}{\gamma}{\delta}{\varepsilon}{\zeta}{\eta}}&=&
      \frac{1}{4}C_{{\alpha}{\gamma}{\beta}{\delta};{\varepsilon}{\zeta}{\eta}}
      +\frac{1}{4}(C_{{\alpha}{\gamma}}{^{\mu}}{_{\varepsilon}}
         C_{{\beta}{\delta}{\mu}{\zeta}})_{;{\eta}}
      -\frac{1}{5}g_{{\alpha}{\gamma}}(C_{\beta}{^{{\mu}{\nu}}}{_{\varepsilon}}
         C_{{\delta}{\mu}{\nu}{\zeta}})_{;{\eta}}\nonumber\\
      &&-\frac{1}{10}g_{{\alpha}{\gamma}}C_{\beta}{^{{\mu}{\nu}}}_{\varepsilon}
         C_{{\delta}{\mu}{\nu}{\zeta};{\eta}}
      +\frac{1}{15}g_{{\alpha}{\gamma}}C_{{\beta}{\delta}{\mu}{\varepsilon};{\nu}}
         C^{\mu}{_{\zeta}}{^{\nu}}{_{\eta}}
      +\frac{1}{240} {\Pi}_{{\alpha}{\beta}{\gamma}{\delta}} g_{{\varepsilon}{\zeta}}
         C^{{\mu}{\nu}{\rho}{\sigma}}C_{{\mu}{\nu}{\rho}{\sigma};{\eta}},
\label{eq:v04}
\end{eqnarray}
where
\begin{equation}
{\Pi}_{{\alpha}{\beta}{\gamma}{\delta}} = {1 \over 2} g_{{\alpha}{\gamma}}
g_{{\beta}{\delta}} + {1 \over 2} g_{{\alpha}{\delta}} g_{{\beta}{\gamma}} +
{\kappa}g_{{\alpha}{\beta}} g_{{\gamma}{\delta}}.
\end{equation}
In Eqs.\ (\ref{eq:v01}) -- (\ref{eq:v04}), the right hand sides are
understood to be symmetrized on the index par $({\alpha}{\beta})$, on the
index pair $({\gamma}{\delta})$, and on as many of the index triplet
$({\varepsilon}{\zeta}{\eta})$ as are present.
When ${\kappa}=-1/2$, the formulae (\ref{eq:v01}) -- (\ref{eq:v03}) agree
with Eqs.\ (A20) -- (A22) of AFO specialized to the vacuum case.

We will also need the first two of the coefficients in the expansion of
$V^1_{{\alpha}{\beta}{\alpha}'{\beta}'}$, which we obtain from Eq.\
(\ref{eq:recrel2}) with $n=1$.  The hard part of the computation is
evaluating the source term on the right hand side of this equation.   
Using the definition (\ref{eq:CTSEVn}) and the expansions
(\ref{eq:sigmaexpand}), (\ref{eq:Qexpand}) and (\ref{eq:Zexpand}) we
obtain  
\begin{eqnarray}
g^{\ \,{\alpha}'}_{\gamma}g^{\ \,{\beta}'}_{\delta}\, {\Box}
V^0_{{\alpha}{\beta}{\alpha}^{\prime}{\beta}^{\prime}} &=&  {\Box}
v^0_{{\alpha}{\beta}{\gamma}{\delta}} + 2
v_{{\alpha}{\beta}{\gamma}{\delta}{\varepsilon}}^{0\ \ \ \ \ \,;{\varepsilon}}
+ v_{{\alpha}{\beta}{\gamma}{\delta}{\varepsilon}}^{0\ \ \ \ \ \ {\varepsilon}}
\nonumber \\ \mbox{} && 
+ \left[ {\Box}v^0_{{\alpha}{\beta}{\gamma}{\delta}{\varepsilon}} + 2
  v_{{\alpha}{\beta}{\gamma}{\delta}{\varepsilon}{\zeta}}^{0\ \ \ \ \ \ \,;{\zeta}}
+ v_{{\alpha}{\beta}{\gamma}{\delta}{\varepsilon}{\zeta}}^{0\ \ \ \ \ \ \,\,{\zeta}}
- C_{{\alpha}{\eta}{\beta}{\delta};{\zeta}} C_{{\gamma}\ \ \, \, {\varepsilon}}^{\ \,
  {\eta}{\zeta}}   
\right] {\sigma}^{;{\varepsilon}} + {\calO}(s^2).
\label{eq:intermediate1}
\end{eqnarray}
Now inserting suitably symmetrized versions of the 
formulae (\ref{eq:v01}) -- (\ref{eq:v04}) for the expansion
coefficients $v^0_{{\alpha}\ldots{\eta}}$ and using the identity
(\ref{eq:identity4}) gives
\begin{eqnarray}
g^{{\alpha}'}_{\gamma}g^{{\beta}'}_{\delta}\, {\Box}
V^0_{{\alpha}{\beta}{\alpha}^{\prime}{\beta}^{\prime}} &=&  
- \frac{1}{3} {\Box}C_{{\alpha}{\gamma}{\beta}{\delta}} 
- \frac{1}{6} C_{{\alpha}{\gamma}{\eta}{\zeta}} C_{{\beta}{\delta}}^{\ \ \
  {\eta}{\zeta}}
+ \frac{1}{24} g_{{\alpha}{\gamma}} g_{{\beta}{\delta}}
C_{{\mu}{\nu}{\rho}{\sigma}}C^{{\mu}{\nu}{\rho}{\sigma}}
- \frac{1}{180} {\Pi}_{{\alpha}{\beta}{\gamma}{\delta}}
C_{{\mu}{\nu}{\rho}{\sigma}}C^{{\mu}{\nu}{\rho}{\sigma}} 
\nonumber \\ \mbox{} && 
+ \bigg[ 
\frac{1}{12} {\Box}(C_{{\alpha}{\gamma}{\beta}{\delta};{\varepsilon}})
+ \frac{1}{360} {\Pi}_{{\alpha}{\beta}{\gamma}{\delta}}
C_{{\mu}{\nu}{\rho}{\sigma};{\varepsilon}}C^{{\mu}{\nu}{\rho}{\sigma}} 
- \frac{1}{24} g_{{\alpha}{\gamma}} g_{{\beta}{\delta}} 
C_{{\mu}{\nu}{\rho}{\sigma};{\varepsilon}}C^{{\mu}{\nu}{\rho}{\sigma}} 
+ \frac{1}{6} C_{{\alpha}{\gamma}}^{\ \ \ {\mu}{\nu}}
C_{{\beta}{\delta}{\mu}{\nu};{\varepsilon}}
\nonumber \\ \mbox{} && 
- \frac{1}{6} C_{{\alpha}{\gamma}}^{\ \ \ {\mu}{\nu}}
C_{{\beta}{\delta}{\mu}{\varepsilon};{\nu}}
+ \frac{5}{6} C_{{\alpha}{\eta}{\beta}{\delta};{\zeta}} C^{\ {\zeta}\ \
  {\eta}}_{{\varepsilon}\ \, {\gamma}}
+ \frac{1}{6} C_{{\gamma}{\eta}{\beta}{\delta};{\zeta}} C^{\ {\zeta}\ \
  {\eta}}_{{\varepsilon}\ \, {\alpha}}
+ \frac{1}{45} g_{{\beta}{\delta}} C_{{\alpha}{\gamma}{\eta}{\zeta};{\rho}}
C^{{\eta}{\zeta}{\rho}}_{\ \ \ \ \,{\varepsilon}} 
\nonumber \\ \mbox{} && 
+ \frac{1}{15} g_{{\beta}{\delta}} C_{\alpha}^{\ \ {\eta}{\zeta}{\rho}}
C_{{\gamma}{\eta}{\varepsilon}{\rho};{\zeta}}
+ \frac{1}{45} g_{{\beta}{\delta}} C_{{\alpha}{\gamma}{\eta}{\zeta};{\rho}}
C^{{\rho}{\zeta}{\eta}}_{\ \ \ \ \,{\varepsilon}} 
+ \frac{1}{10} g_{{\beta}{\delta}} C_{\gamma}^{\ \ {\eta}{\zeta}{\rho}}
C_{{\alpha}{\eta}{\varepsilon}{\rho};{\zeta}}
\bigg] {\sigma}^{;{\varepsilon}} + {\calO}(s^2),
\label{eq:intermediate2}
\end{eqnarray}
where the right hand side is understood to be symmetrized on the index
pairs $({\alpha}{\beta})$ and $({\gamma}{\delta})$.  
Next, we expand both sides of Eq.\ (\ref{eq:recrel2}) with $n=1$ as a
power series in ${\sigma}_{;{\mu}}$ to ${\calO}(s)$, using the
formulae (\ref{eq:v01}), (\ref{eq:v02}) and (\ref{eq:intermediate2})
to evaluate the right hand side, and the expansion (\ref{eq:CTSEVn})
with $n=1$ on the left hand side.  
Equating the coefficients of the various powers of ${\sigma}_{;{\mu}}$
then gives a series of equations that can be solved for the
coefficients $v^1_{{\alpha}\ldots{\eta}}$.  The results are
\begin{eqnarray}
   v^1_{{\alpha}{\beta}{\gamma}{\delta}}&=&
      \frac{1}{12}{\Box}C_{{\alpha}{\gamma}{\beta}{\delta}}
      +\frac{1}{2}C_{\alpha}{^{\mu}}{_{\beta}}{^{\nu}}C_{{\gamma}{\mu}{\delta}{\nu}}
      +\frac{1}{24}C_{{\alpha}{\gamma}}{^{{\mu}{\nu}}}C_{{\beta}{\delta}{\mu}{\nu}}
      -\frac{1}{24}g_{{\alpha}{\gamma}}C_{\beta}{^{{\mu}{\nu}{\rho}}}C_{{\delta}{\mu}{\nu}{\rho}}
      +\frac{1}{720}{\Pi}_{{\alpha}{\beta}{\gamma}{\delta}} C^{{\mu}{\nu}{\rho}{\sigma}}
         C_{{\mu}{\nu}{\rho}{\sigma}}
\label{eq:v11}
\end{eqnarray}
and
\begin{eqnarray}
   v^1_{{\alpha}{\beta}{\gamma}{\delta}{\varepsilon}}&=&
-\frac{1}{24} {\Box}(C_{{\alpha}{\gamma}{\beta}{\delta};{\varepsilon}})
- \frac{1}{720} {\Pi}_{{\alpha}{\beta}{\gamma}{\delta}}
C_{{\mu}{\nu}{\rho}{\sigma};{\varepsilon}}C^{{\mu}{\nu}{\rho}{\sigma}} 
+ \frac{1}{80} g_{{\alpha}{\gamma}} g_{{\beta}{\delta}} 
C_{{\mu}{\nu}{\rho}{\sigma};{\varepsilon}}C^{{\mu}{\nu}{\rho}{\sigma}} 
- \frac{1}{18} C_{{\alpha}{\gamma}}^{\ \ \ {\mu}{\nu}}
C_{{\beta}{\delta}{\mu}{\nu};{\varepsilon}}
\nonumber \\ \mbox{} && 
- \frac{1}{3} C_{{\alpha}\ \, {\beta}}^{\ \, {\mu}\ \, {\nu}}
C_{{\mu}{\gamma}{\nu}{\delta};{\varepsilon}} 
- \frac{1}{6} C_{{\alpha}\ \, {\beta}\ \, ;{\varepsilon}}^{\ \, {\mu}\ \, {\nu}}
C_{{\mu}{\gamma}{\nu}{\delta}} 
- \frac{1}{12} C_{{\varepsilon}\ \, {\alpha}}^{\ \, {\eta}\ \, {\zeta}}
C_{{\zeta}{\gamma}{\beta}{\delta};{\eta}} 
- \frac{1}{4} C_{{\varepsilon}{\eta}{\gamma}}^{\ \ \ \ {\zeta}}
C_{{\alpha}{\zeta}{\beta}{\delta}}^{\ \ \ \ \ \,;{\eta}}
+ \frac{1}{36} C_{{\alpha}{\gamma}{\zeta}{\eta}} C_{{\beta}{\delta}\ \ 
  {\varepsilon}}^{\ \ \ {\zeta}\ \, ;{\eta}}
\nonumber \\ \mbox{} && 
- \frac{1}{270} g_{{\beta}{\delta}} C_{{\alpha}{\gamma}{\eta}{\zeta};{\rho}}
C^{{\eta}{\zeta}{\rho}}_{\ \ \ \ \,{\varepsilon}} 
- \frac{1}{90} g_{{\beta}{\delta}} C_{\alpha}^{\ \ {\eta}{\zeta}{\rho}}
C_{{\gamma}{\eta}{\varepsilon}{\rho};{\zeta}}
- \frac{1}{270} g_{{\beta}{\delta}} C_{{\alpha}{\gamma}{\eta}{\zeta};{\rho}}
C^{{\rho}{\zeta}{\eta}}_{\ \ \ \ \,{\varepsilon}} 
- \frac{1}{60} g_{{\beta}{\delta}} C_{\gamma}^{\ \ {\eta}{\zeta}{\rho}}
C_{{\alpha}{\eta}{\varepsilon}{\rho};{\zeta}} \nonumber \\
\mbox{} && + \frac{1}{180} g_{{\beta}{\delta}} C_{\alpha}^{\ \,{\mu}{\nu}{\rho}}
C_{{\gamma}{\mu}{\nu}{\rho};{\varepsilon}},
\label{eq:v12}
\end{eqnarray}
where again there is implicit symmetrization on the index pairs
(${\alpha}{\beta}$) and (${\gamma}{\delta}$).
When ${\kappa}=-1/2$, Eq. (\ref{eq:v11}) agrees
with Eq.\ (A23) of AFO specialized to the vacuum case\footnote{Note that the
expressions (\protect{\ref{eq:v01}}) -- (\protect{\ref{eq:v04}})
and (\ref{eq:v11}) -- (\ref{eq:v12})
are all traceless on the index pair $({\gamma}{\delta})$, aside from the
terms involving the  
tensor ${\Pi}_{{\alpha}{\beta}{\gamma}{\delta}}$.  This means that performing
a trace reversal on the index pair $({\gamma}{\delta})$ is equivalent to
changing the value of ${\kappa}$ from $0$ to $-1/2$, in agreement with
the discussion after Eq.\ (\protect{\ref{eq:LGFE0}}) above.}.

\section{Expansion coefficients}
\label{sec:Vcoeffs}

In this appendix we list the expressions for the coefficients ${\cal
V}_{{\alpha}\ldots {\eta}}$ which appear in the expansion
(\ref{eq:mainexpand}).   These expressions are obtained by
substituting Eqs.\ (\protect{\ref{eq:v01}}) -- (\protect{\ref{eq:v04}})
and (\ref{eq:v11}) -- (\ref{eq:v12}), specialized to ${\kappa}=0$, into
Eqs.\ (\ref{eq:calVdef0}) -- (\ref{eq:calVdef2}).  The results are
\begin{equation}
\label{eq:calV0ans}
{\cal V}_{{\alpha}{\beta}{\gamma}{\delta}{\varepsilon}} = - \frac{1}{2}
C_{{\alpha}{\gamma}{\beta}{\delta};{\varepsilon}},
\end{equation}
\begin{eqnarray}
\label{eq:calV1ans}
{\cal V}_{{\alpha}{\beta}{\gamma}{\delta}{\varepsilon}}^{\ \ \ \ \ \ \ {\sigma}}
&=&  \frac{1}{6} C_{{\alpha}{\gamma}{\beta}{\delta};{\varepsilon}}^{\ \ \ \ \ \ \
  \, ;{\sigma}},
+ \frac{2}{3} C_{{\varepsilon}\ \, {\alpha}}^{\ \, {\sigma}\ \, {\rho}}
C_{{\rho}{\gamma}{\beta}{\delta}} 
- \frac{1}{3} C_{{\varepsilon}\ \, {\gamma}}^{\ \, {\sigma}\ \, {\rho}}
C_{{\alpha}{\rho}{\beta}{\delta}} 
- \frac{1}{6} C_{{\alpha}{\gamma}\ \,{\varepsilon}}^{\ \ \ {\rho}}
C_{{\beta}{\gamma}{\rho}}^{\ \ \ \ {\sigma}}
+ \frac{1}{12} g_{{\alpha}{\gamma}} C_{{\beta}\ \ \ {\varepsilon}}^{\
  \,{\rho}{\tau}}
C_{{\delta}{\rho}{\tau}}^{\ \ \ \ {\sigma}} \nonumber \\
\mbox{} && 
+ \frac{1}{12} g_{{\alpha}{\gamma}} C_{\beta}^{\
  \,{\rho}{\tau}{\sigma}} C_{{\delta}{\rho}{\tau}{\varepsilon}} 
+ {\delta}_{\varepsilon}^{\sigma}\bigg[ 
\frac{1}{12} {\Box}C_{{\alpha}{\gamma}{\beta}{\delta}}
+ \frac{1}{2} C_{{\alpha}\,\ {\beta}}^{\ \, {\rho}\ \, {\tau}} 
C_{{\gamma}{\rho}{\delta}{\tau}}
+ \frac{1}{24} C_{{\alpha}{\gamma}}^{\ \ \ {\rho}{\tau}} C_{{\beta}{\delta}{\rho}{\tau}}
- \frac{1}{96} g_{{\alpha}{\gamma}} g_{{\beta}{\delta}} C_{{\rho}{\tau}{\mu}{\nu}}
C^{{\rho}{\tau}{\mu}{\nu}}\bigg],
\nonumber \\
\end{eqnarray}
and
\begin{eqnarray}
\label{eq:calV2ans}
{\cal V}_{{\alpha}{\beta}({\gamma}{\delta}|{\varepsilon}|{\sigma}{\rho})}
&=& - \frac{1}{24} C_{{\alpha}{\gamma}{\beta}{\delta};{\varepsilon}{\sigma}{\rho}}
+ \frac{1}{720} g_{{\alpha}{\gamma}} g_{{\beta}{\delta}} g_{{\varepsilon}{\sigma}}
C_{{\tau}{\lambda}{\mu}{\nu}} C^{{\tau}{\lambda}{\mu}{\nu}}_{\ \ \ \ \ \ ;{\rho}}
+ \frac{1}{12} C_{{\alpha}{\gamma}{\tau}{\delta}} C_{{\beta}{\sigma}\
  \,{\varepsilon};{\rho}}^{\ \ \ {\tau}}
- \frac{1}{6} C_{{\alpha}{\tau}{\varepsilon}{\gamma};{\delta}} C^{\tau}_{\
  \,{\sigma}{\beta}{\rho}} 
+ \frac{1}{6} C_{{\alpha}{\tau}{\beta}{\gamma}} C_{{\varepsilon}{\delta}{\sigma}\ \,
  ;{\rho}}^{\ \ \ \,\, {\tau}}
\nonumber \\ \mbox{} && 
+ \frac{1}{12} C_{{\alpha}{\gamma}{\tau}{\varepsilon}} C_{{\beta}{\delta}\
  \,{\sigma};{\rho}}^{\ \ \, {\tau}}
- \frac{5}{12} C_{{\alpha}{\tau}{\varepsilon}{\gamma}} C^{\tau}_{\
  \,{\delta}{\beta}{\sigma};{\rho}} 
+ \frac{1}{12} C_{{\alpha}{\tau}{\beta}{\gamma};{\delta}}
C_{{\varepsilon}{\sigma}{\rho}}^{\ \ \ \ {\tau}}
+ \frac{1}{24} C_{{\alpha}{\gamma}{\beta}{\delta};{\tau}}
C_{{\varepsilon}{\sigma}{\rho}}^{\ \ \ \ {\tau}}
- \frac{1}{12} C_{{\alpha}{\gamma}{\tau}{\delta}} C_{{\beta}{\sigma}\
  \,{\rho};{\varepsilon}}^{\ \ \ {\tau}}
\nonumber \\ \mbox{} && 
- g_{{\beta}{\gamma}} \bigg[ \frac{1}{90}
  C_{{\delta}{\alpha}{\tau}{\varepsilon};{\lambda}}
C_{{\sigma}\ \,{\rho}}^{\ \,{\tau}\ \, {\lambda}}
+ \frac{1}{30} C_{{\delta}{\tau}{\sigma}{\lambda}} C_{{\alpha}\ \,{\varepsilon}\
  \,;{\rho}}^{\ \,{\tau}\ \,{\lambda}}
+ \frac{1}{20} C_{{\delta}{\tau}{\varepsilon}{\lambda};{\sigma}} C_{{\alpha}\ \,
  {\rho}}^{\ \,{\tau}\, \ {\lambda}}
+ \frac{1}{90} C_{{\delta}{\alpha}{\tau}{\sigma};{\lambda}} 
C_{{\varepsilon}\ \,{\rho}}^{\ \,{\tau}\ \,{\lambda}}
+ \frac{1}{90} C_{{\delta}{\alpha}{\tau}{\sigma};{\lambda}} 
C_{{\rho}\ \,{\varepsilon}}^{\ \,{\tau}\ \,{\lambda}}
\nonumber \\ \mbox{} && 
+ \frac{1}{30} C_{{\delta}{\tau}{\varepsilon}{\lambda}} 
C_{{\alpha}\ \,{\sigma}\ \,;{\rho}}^{\ \, {\tau}\ \,{\lambda}}
+ \frac{1}{20} C_{{\delta}{\tau}{\sigma}{\lambda};{\rho}} C_{{\alpha}\
  \,{\varepsilon}}^{\ \,{\tau}\ \,{\lambda}}
- \frac{1}{20} C_{{\delta}{\tau}{\sigma}{\lambda}} C_{{\alpha}\ \,{\rho}\
  \,;{\varepsilon}}^{\ \,{\tau}\ \,{\lambda}}
- \frac{1}{30} C_{{\delta}{\tau}{\sigma}{\lambda};{\varepsilon}} C_{{\alpha}\
  \,{\rho}}^{\ \,{\tau}\ \,{\lambda}}
\bigg]
\nonumber \\ \mbox{} &&
+ g_{{\varepsilon}{\rho}} \bigg[ 
 \frac{1}{90}  g_{{\alpha}{\gamma}} g_{{\beta}{\delta}}
  C_{{\tau}{\lambda}{\mu}{\nu};{\sigma}} C^{{\tau}{\lambda}{\mu}{\nu}}
- \frac{1}{24} {\Box}( C_{{\alpha}{\gamma}{\beta}{\delta};{\sigma}})
- \frac{1}{6} C_{{\alpha}{\tau}{\beta}{\lambda};{\gamma}} C^{{\tau}\ \,{\lambda}}_{\
  \,{\delta}\ \, {\sigma}}
- \frac{1}{18} C_{{\alpha}{\gamma}{\tau}{\lambda};{\delta}} C_{{\beta}{\sigma}}^{\ \
  \ {\tau}{\lambda}}
- \frac{1}{4} C_{{\gamma}{\tau}{\delta}{\lambda}}
C_{{\alpha}\ \,{\beta}{\sigma}}^{\ \,{\lambda}\ \ \ ;{\tau}}
\nonumber \\ \mbox{} &&
- \frac{1}{12} C_{{\gamma}{\tau}{\alpha}{\lambda}} C^{{\lambda}\ \ \ \
  ;{\tau}}_{\ \,{\delta}{\beta}{\sigma}}
+ \frac{1}{36} C_{{\alpha}{\gamma}{\tau}{\delta};{\lambda}} C_{{\beta}{\sigma}}^{\ \
  \ {\tau}{\lambda}}
+ \frac{1}{270} g_{{\beta}{\gamma}} C_{{\delta}{\alpha}{\tau}{\lambda};{\mu}}
C^{{\lambda}{\tau}\ \,{\mu}}_{\ \ \ {\sigma}}
- \frac{1}{90} g_{{\beta}{\gamma}} C_{{\delta}{\tau}{\sigma}{\lambda};{\mu}}
C_{\alpha}^{\ \,{\tau}{\mu}{\lambda}}
\nonumber \\ \mbox{} &&
- \frac{1}{60} g_{{\beta}{\gamma}} C_{{\delta}{\tau}{\lambda}{\mu}} C_{{\alpha}\
  \,{\sigma}}^{\ \, {\tau}\ \,\,{\mu};{\lambda}}
+ \frac{1}{270} g_{{\beta}{\gamma}} C_{{\gamma}{\alpha}{\tau}{\lambda};{\mu}}
C_{\sigma}^{\ \,{\tau}{\lambda}{\mu}}
+ \frac{1}{180} g_{{\beta}{\gamma}} C_{\alpha}^{\ \,{\tau}{\lambda}{\mu}}
C_{{\delta}{\tau}{\lambda}{\mu};{\sigma}}  
- \frac{1}{3} C_{{\alpha}{\tau}{\beta}{\lambda}} C_{{\gamma}\ \, {\delta}\ \,\,
  ;{\sigma}}^{\ \, {\tau}\ \,{\lambda}} \bigg]
\nonumber \\ \mbox{} &&
+ g_{{\sigma}{\rho}} \bigg[ 
- \frac{1}{240} g_{{\alpha}{\gamma}} g_{{\beta}{\delta}} 
C_{{\tau}{\lambda}{\mu}{\nu}} C^{{\tau}{\lambda}{\mu}{\nu}}_{\ \ \ \ \ \ ;{\varepsilon}}
+ \frac{1}{48} {\Box}( C_{{\alpha}{\gamma}{\beta}{\delta};{\varepsilon}})
+ \frac{1}{24} C_{{\alpha}{\tau}{\beta}{\gamma};{\lambda}} C_{{\varepsilon}\
  \,{\delta}}^{\ \,{\lambda}\ \,{\tau}}
+ \frac{1}{8} C_{{\varepsilon}{\tau}{\alpha}{\lambda}} C^{{\lambda}\ \ \ \
  ;{\tau}}_{\ \,{\gamma}{\beta}{\delta}}
\nonumber \\ \mbox{} &&
+ \frac{1}{72} C_{{\alpha}{\gamma}{\tau}{\varepsilon};{\lambda}}
C_{{\beta}{\delta}}^{\ \ \ {\tau}{\lambda}}
+ \frac{1}{6} C_{{\alpha}{\tau}{\beta}{\lambda};{\varepsilon}} C_{{\gamma}\,\
  {\delta}}^{\ \,{\tau}\ \,{\lambda}}
+ \frac{1}{72} C_{{\alpha}{\gamma}{\tau}{\lambda};{\varepsilon}}
C_{{\beta}{\delta}}^{\ \ \ {\tau}{\lambda}}
+ \frac{1}{12} C_{{\alpha}{\tau}{\beta}{\lambda}} C_{{\gamma}\,\
  {\delta}\ \,\,;{\varepsilon}}^{\ \,{\tau}\ \,{\lambda}}
+ \frac{1}{540} g_{{\beta}{\gamma}} C_{{\delta}{\alpha}{\tau}{\lambda};{\mu}}
C^{{\lambda}{\tau}\ \,{\mu}}_{\ \ \ {\varepsilon}}
\nonumber \\ \mbox{} &&
- \frac{1}{180} g_{{\beta}{\gamma}} C_{{\delta}{\tau}{\varepsilon}{\lambda};{\mu}}
C_{\alpha}^{\ \,{\tau}{\mu}{\lambda}}
- \frac{1}{120} g_{{\beta}{\gamma}} C_{{\delta}{\tau}{\lambda}{\mu}} C_{{\alpha}\
  \,{\varepsilon}} ^{\ \,{\tau}\ \,{\mu};{\lambda}}
+ \frac{1}{540} g_{{\beta}{\gamma}} C_{{\delta}{\alpha}{\tau}{\lambda};{\mu}}
C_{\varepsilon}^{\ \,{\tau}{\lambda}{\mu}}
+ \frac{1}{360} g_{{\beta}{\gamma}} C_{{\alpha}{\tau}{\lambda}{\mu}} C_{{\delta}\ \
  \ \ \ ;{\varepsilon}}^{\ \, {\tau}{\lambda}{\mu}}
\bigg].
\end{eqnarray}
The right hand sides of Eqs.\ (\ref{eq:calV0ans}) and
(\ref{eq:calV1ans}) are understood to be symmetrized over the index
pairs $({\alpha}{\beta})$ and $({\gamma}{\delta})$.  In Eq.\
(\ref{eq:calV2ans}), as the notation on the left hand side indicates,
we have only computed the piece of ${\cal
V}_{{\alpha}{\beta}{\gamma}{\delta}{\varepsilon}{\sigma}{\rho}}$ which is totally
symmetric on the indices $({\gamma}{\delta}{\sigma}{\rho})$.  This is because
only this piece of ${\cal
V}_{{\alpha}{\beta}{\gamma}{\delta}{\varepsilon}{\sigma}{\rho}}$ is needed for
computing the expansion coefficient $f^{(2)}_{\alpha}$, from Eq.\
(\ref{eq:f3ans1}) above.  The right hand side of Eq.\
(\ref{eq:calV2ans}) is understood to be symmetrized on the index
pair $({\alpha}{\beta})$ and on the indices $({\gamma}{\delta}{\sigma}{\rho})$.

\section{Some Useful Determinant Identities}
\label{s:DetIDs}
One occasionally encounters determinants of rank two tensors in relativity.
The normal expressions for evaluating determinants do not lend themselves
naturally to expression using standard tensor notation. However, it is
possible to write the determinant of any square matrix in terms of powers of
the matrix and their traces, both of which are easily expressed in tensor
notation. We list the three simplest here for a matrix $A$ of various sizes
\begin{eqnarray}
   \mbox{\textit{Dimension}}&&\mbox{\textit{Identity}} \nonumber\\
   2{\times}2&&Det\,A = \left[ (\mbox{\textit{Tr}} A )^2 - \mbox{\textit{Tr}} (A^2)
      \right]/2 \label{e:Det2}\\
   3{\times}3&&Det\,A = \left[ (\mbox{\textit{Tr}} A )^3 - 3 (\mbox{\textit{Tr}}
      A)\mbox{\textit{Tr}} (A^2) + 2 \mbox{\textit{Tr}}(A^3) \right]/6
      \label{e:Det3}\\
      4{\times}4&&Det\,A = \left[ (\mbox{\textit{Tr}} A )^4 - 6 (\mbox{\textit{Tr}}
         A)^2\mbox{\textit{Tr}} (A^2) + 8 (\mbox{\textit{Tr}} A)
         \mbox{\textit{Tr}}(A^3) + 3 (\mbox{\textit{Tr}}(A^2))^2 - 6
         \mbox{\textit{Tr}}(A^4) \right]/24 \label{e:Det4}
\end{eqnarray}
Similar identities can be obtained for any dimension through a straightforward
recursive application of Newton's Identities. 

To make the usefulness of these identities more apparent, we express the
determinant in the $4{\times}4$ case in tensor notation using (\ref{e:Det4}):
\begin{equation}
   Det\,A^{\alpha}{_{\beta}} = \left[\left(A^{\alpha}{_{\alpha}}\right)^4 - 6 \left(A^{\alpha}{_{\alpha}}\right)^2
   A^{\beta}{_{\gamma}}A^{\gamma}{_{\beta}} + 8 A^{\alpha}{_{\alpha}} A^{\beta}{_{\gamma}}A^{\gamma}{_{\delta}}A^{\delta}{_{\beta}} + 3
   \left(A^{\alpha}{_{\beta}}A^{\beta}{_{\gamma}}\right)^2-6 A^{\alpha}{_{\beta}}A^{\beta}{_{\gamma}}A^{\gamma}{_{\delta}}A^{\delta}{_{\alpha}}\right]/24.
   \label{e:DetTr}
\end{equation}
While this expression is not a computationally efficient way of calculating
the determinant, it clearly has the advantage of reducing the determinant
operation to more familiar tensor operations. We include it here because we
have not been successful in locating it in the literature.


\begin{thebibliography}{8}
\bibitem{Poisson03} E. Poisson, {\it The motion of point particles in
  curved spacetime}, gr-qc/0306052.  Note that the sign convention for
  $V$ in this review is the opposite from that used here.
\end{thebibliography}
\end{document}